
%
\def\unredoffs{}
\tolerance=1000\hfuzz=2pt
\catcode`\@=11 
\ifx\hyperdef\UNd@FiNeD\def\hyperdef#1#2#3#4{#4}\def\hyperref#1#2#3#4{#4}\def\href#1#2{#2}\fi
\magnification=1200\unredoffs\baselineskip=16pt plus 2pt minus 1pt
\def\Date#1{\vfill\leftline{#1}\tenpoint\supereject%
\footline={\hss\tenrm\hyperdef\hypernoname{page}\folio\folio\hss}}%

{\count255=\time\divide\count255 by 60 \xdef\hourmin{\number\count255}
 \multiply\count255 by-60\advance\count255 by\time
 \xdef\hourmin{\hourmin:\ifnum\count255<10 0\fi\the\count255}
}
\def\date{\number\day.\number\month.\number\year\ at \hourmin}


\def\nolabels{\def\wrlabeL##1{}\def\eqlabeL##1{}\def\reflabeL##1{}}
\def\writelabels{\def\wrlabeL##1{\leavevmode\vadjust{\rlap{\smash%
{\line{{\escapechar=` \hfill\rlap{\sevenrm\hskip.03in\string##1}}}}}}}%
\def\eqlabeL##1{{\escapechar-1\rlap{\sevenrm\hskip.05in\string##1}}}%
\def\reflabeL##1{\noexpand\llap{\noexpand\sevenrm\string\string\string##1}}}
\nolabels

\global\newcount\secno \global\secno=0
\global\newcount\meqno \global\meqno=1
\def\s@csym{}

\def\newsec#1\par{\global\advance\secno by1%
{\toks0{#1}\message{(\the\secno. \the\toks0)}}%
\global\subsecno=0\eqnres@t\let\s@csym\secsym\xdef\secn@m{\the\secno}\noindent
{\bf\hyperdef\hypernoname{section}{\the\secno}{\the\secno.} #1}%
\writetoca{{\string\hyperref{}{section}{\the\secno}{\bf \the\secno\quad}} {\bf #1}}\par%
\nobreak\medskip\nobreak\noindent\ignorespaces}
\def\eqnres@t{\xdef\secsym{\the\secno.}\global\meqno=1\bigbreak\bigskip}
\def\sequentialequations{\def\eqnres@t{\bigbreak}}\xdef\secsym{}

\global\newcount\subsecno \global\subsecno=0
\def\subsec#1\par{\global\advance\subsecno by1%
{\toks0{#1}\message{(\s@csym\the\subsecno. \the\toks0)}}%
\global\subsubsecno=0%
\ifnum\lastpenalty>9000\else\bigbreak\fi
\noindent{\it\hyperdef\hypernoname{subsection}{\secn@m.\the\subsecno}%
{\secn@m.\the\subsecno.} #1}\writetoca{\string\hskip1.45cm
{\string\hyperref{}{subsection}{\secn@m.\the\subsecno}{\secn@m.\the\subsecno.}}
{#1}}\par\nobreak\medskip\nobreak\noindent\ignorespaces}

\def\appendix#1#2{\global\meqno=1\global\subsecno=0\xdef\secsym{\hbox{#1.}}%
\bigbreak\bigskip\noindent{\bf Appendix \hyperdef\hypernoname{appendix}{#1}%
{#1.} #2}{\toks0{(#1. #2)}\message{\the\toks0}}%
\xdef\s@csym{#1.}\xdef\secn@m{#1}%
\writetoca{{\string\hyperref{}{appendix}{#1}{\bf {#1}\quad}} {\bf #2}}%
\par\nobreak\medskip\nobreak}

%
\def\checkm@de#1#2{\ifmmode{\def\f@rst##1{##1}\hyperdef\hypernoname{equation}%
{#1}{#2}}\else\hyperref{}{equation}{#1}{#2}\fi}
\def\eqnn#1{\DefWarn#1\xdef #1{(\noexpand\relax\noexpand\checkm@de%
{\s@csym\the\meqno}{\secsym\the\meqno})}%
\wrlabeL#1\writedef{#1\leftbracket#1}\global\advance\meqno by1}
\def\f@rst#1{\c@t#1a\em@ark}\def\c@t#1#2\em@ark{#1}
\def\eqna#1{\DefWarn#1\wrlabeL{#1$\{\}$}%
\xdef #1##1{(\noexpand\relax\noexpand\checkm@de%
{\s@csym\the\meqno\noexpand\f@rst{##1}1}{\hbox{$\secsym\the\meqno##1$}})}
\writedef{#1\numbersign1\leftbracket#1{\numbersign1}}\global\advance\meqno by1}
\def\eqn#1#2{\DefWarn#1%
\xdef #1{(\noexpand\hyperref{}{equation}{\s@csym\the\meqno}%
{\secsym\the\meqno})}$$#2\eqno(\hyperdef\hypernoname{equation}%
{\s@csym\the\meqno}{\secsym\the\meqno})\eqlabeL#1$$%
\writedef{#1\leftbracket#1}\global\advance\meqno by1}
\def\xeqn{\expandafter\xe@n}\def\xe@n(#1){#1}
\def\xeqna#1{\expandafter\xe@n#1}
\def\eqns#1{(\e@ns #1{\hbox{}})}
\def\e@ns#1{\ifx\UNd@FiNeD#1\message{eqnlabel \string#1 is undefined.}%
\xdef#1{(?.?)}\fi{\let\hyperref=\relax\xdef\next{#1}}%
\ifx\next\em@rk\def\next{}\else%
\ifx\next#1\xeqn#1\else\def\n@xt{#1}\ifx\n@xt\next#1\else\xeqna#1\fi
\fi\let\next=\e@ns\fi\next}

\def\DefWarn#1{\ifx\UNd@FiNeD#1\else
\immediate\write16{*** WARNING: the label \string#1 is already defined ***}\fi}
%
\newskip\footskip\footskip14pt plus 1pt minus 1pt 
\def\footnotefont{\ninepoint}\def\f@t#1{\footnotefont #1\@foot}
\def\f@@t{\baselineskip\footskip\bgroup\footnotefont\aftergroup\@foot\let\next}
\setbox\strutbox=\hbox{\vrule height9.5pt depth4.5pt width0pt}
\global\newcount\ftno \global\ftno=0
\def\foot{\global\advance\ftno by1\def\foot@rg{\hyperref{}{footnote}%
{\the\ftno}{\the\ftno}\xdef\foot@rg{\noexpand\hyperdef\noexpand\hypernoname%
{footnote}{\the\ftno}{\the\ftno}}}\footnote{$^{\foot@rg}$}}
%
%
%
\global\newcount\refno \global\refno=1
\newwrite\rfile
\def\ref{[\hyperref{}{reference}{\the\refno}{\the\refno}]\nref}
\def\nref#1{\DefWarn#1%
\xdef#1{[\noexpand\hyperref{}{reference}{\the\refno}{\the\refno}]}%
\writedef{#1\leftbracket#1}%
\ifnum\refno=1\immediate\openout\rfile=\jobname.refs\fi
\chardef\wfile=\rfile\immediate\write\rfile{\noexpand\item{[\noexpand\hyperdef%
\noexpand\hypernoname{reference}{\the\refno}{\the\refno}]\ }%
\reflabeL{#1\hskip.31in}\pctsign}\global\advance\refno by1\findarg}
\def\findarg#1#{\begingroup\obeylines\newlinechar=`\^^M\pass@rg}
{\obeylines\gdef\pass@rg#1{\writ@line\relax #1^^M\hbox{}^^M}%
\gdef\writ@line#1^^M{\expandafter\toks0\expandafter{\striprel@x #1}%
\edef\next{\the\toks0}\ifx\next\em@rk\let\next=\endgroup\else\ifx\next\empty%
\else\immediate\write\wfile{\the\toks0}\fi\let\next=\writ@line\fi\next\relax}}
\def\striprel@x#1{} \def\em@rk{\hbox{}}
\def\lref{\begingroup\obeylines\lr@f}
\def\lr@f#1#2{\DefWarn#1\gdef#1{\let#1=\UNd@FiNeD\ref#1{#2}}\endgroup\unskip}
\def\semi{;\hfil\break}
\def\addref#1{\immediate\write\rfile{\noexpand\item{}#1}} 
\def\listrefs{\vfill\supereject\immediate\closeout\rfile\writestoppt
\baselineskip=\footskip\centerline{{\bf References}}\bigskip{\parindent=20pt%
\frenchspacing\escapechar=` \input \jobname.refs\vfill\eject}\nonfrenchspacing}
\def\startrefs#1{\immediate\openout\rfile=\jobname.refs\refno=#1}
\def\xref{\expandafter\xr@f}\def\xr@f[#1]{#1}
\def\refs#1{\count255=1[\r@fs #1{\hbox{}}]}
\def\r@fs#1{\ifx\UNd@FiNeD#1\message{reflabel \string#1 is undefined.}%
\nref#1{need to supply reference \string#1.}\fi%
\vphantom{\hphantom{#1}}{\let\hyperref=\relax\xdef\next{#1}}%
\ifx\next\em@rk\def\next{}%
\else\ifx\next#1\ifodd\count255\relax\xref#1\count255=0\fi%
\else#1\count255=1\fi\let\next=\r@fs\fi\next}
%

%
\newwrite\ffile\global\newcount\figno \global\figno=1
\def\fig{fig.~\hyperref{}{figure}{\the\figno}{\the\figno}\nfig}
\def\nfig#1{\DefWarn#1%
\xdef#1{fig.~\noexpand\hyperref{}{figure}{\the\figno}{\the\figno}}%
\writedef{#1\leftbracket fig.\noexpand~\xfig#1}%
\ifnum\figno=1\immediate\openout\ffile=\jobname.figs\fi\chardef\wfile=\ffile%
{\let\hyperref=\relax
\immediate\write\ffile{\noexpand\medskip\noexpand\item{Fig.\ %
\noexpand\hyperdef\noexpand\hypernoname{figure}{\the\figno}{\the\figno}. }
\reflabeL{#1\hskip.55in}\pctsign}}\global\advance\figno by1\findarg}
\def\xfig{\expandafter\xf@g}\def\xf@g fig.\penalty\@M\ {}
\def\figs#1{figs.~\f@gs #1{\hbox{}}}
\def\f@gs#1{{\let\hyperref=\relax\xdef\next{#1}}\ifx\next\em@rk\def\next{}\else
\ifx\next#1\xfig #1\else#1\fi\let\next=\f@gs\fi\next}
%
\def\figin{\epsfcheck\figin}\def\figins{\epsfcheck\figins}
\def\epsfcheck{\ifx\epsfbox\UnDeFiNeD
\message{(NO epsf.tex, FIGURES WILL BE IGNORED)}
\gdef\figin##1{\vskip2in}\gdef\figins##1{\hskip.5in}
\else\message{(FIGURES WILL BE INCLUDED)}%
\gdef\figin##1{##1}\gdef\figins##1{##1}\fi}
\def\DefWarn#1{}
\def\figinsert{\goodbreak\topinsert}
\def\ifig#1#2#3{\DefWarn#1\xdef#1{fig.~\the\figno}
\writedef{#1\leftbracket fig.\noexpand~\the\figno}%
\figinsert\figin{\centerline{#3}}
\smallskip
\leftskip=0pt \rightskip=0pt
\baselineskip12pt\noindent
{{\bf Fig.~\the\figno}\ \ninepoint #2}
\medskip
\global\advance\figno by1\par\endinsert}
\newwrite\lfile
{\escapechar-1\xdef\pctsign{\string\%}\xdef\leftbracket{\string\{}
\xdef\rightbracket{\string\}}\xdef\numbersign{\string\#}}
\def\writedefs{\immediate\openout\lfile=label.defs \def\writedef##1{%
{\let\hyperref=\relax\let\hyperdef=\relax\let\hypernoname=\relax
 \immediate\write\lfile{\string\def\string##1\rightbracket}}}}%
\def\writestop{\def\writestoppt{\immediate\write\lfile{\string\pageno
 \the\pageno\string\startrefs\leftbracket\the\refno\rightbracket
 \string\def\string\secsym\leftbracket\secsym\rightbracket
 \string\secno\the\secno\string\meqno\the\meqno}\immediate\closeout\lfile}}
\def\writestoppt{}\def\writedef#1{}

\def\seclab#1{\DefWarn#1%
\xdef #1{\noexpand\hyperref{}{section}{\the\secno}{\the\secno}}%
\writedef{#1\leftbracket#1}\wrlabeL{#1=#1}}
\def\subseclab#1{\DefWarn#1%
\xdef #1{\noexpand\hyperref{}{subsection}{\the\secno.\the\subsecno}%
{\the\secno.\the\subsecno}}\writedef{#1\leftbracket#1}\wrlabeL{#1=#1}}
\def\applab#1{\DefWarn#1%
\xdef #1{\noexpand\hyperref{}{appendix}{\secn@m}{\secn@m}}%
\writedef{#1\leftbracket#1}\wrlabeL{#1=#1}}
\newwrite\tfile \def\writetoca#1{}
\def\leaderfill{\leaders\hbox to 1em{\hss.\hss}\hfill}
\def\writetoc{\immediate\openout\tfile=\jobname.toc
   \def\writetoca##1{{\edef\next{\write\tfile{\noindent ##1
   \string\leaderfill{
   \string\hyperref{}{page}{\noexpand\number\pageno}%
   {\noexpand\number\pageno}} \par}}\next}}
}
\newread\ch@ckfile
\def\listtoc{\immediate\closeout\tfile\immediate\openin\ch@ckfile=\jobname.toc
\ifeof\ch@ckfile\message{no file \jobname.toc, no table of contents this pass}%
\else\closein\ch@ckfile\centerline{\bf Contents}\nobreak\medskip%
{\baselineskip=16pt\footnotefont\parskip=0pt\catcode`\@=11\input\jobname.toc
\catcode`\@=12\bigbreak\bigskip}\fi}
\catcode`\@=12 
\def\tenpoint{\def\rm{\fam0\tenrm}
\textfont0=\tenrm \scriptfont0=\sevenrm \scriptscriptfont0=\fiverm
\textfont1=\teni  \scriptfont1=\seveni  \scriptscriptfont1=\fivei
\textfont2=\tensy \scriptfont2=\sevensy \scriptscriptfont2=\fivesy
\textfont\itfam=\tenit \def\it{\fam\itfam\tenit}\def\footnotefont{\ninepoint}%
\textfont\bffam=\tenbf \def\bf{\fam\bffam\tenbf}\def\sl{\fam\slfam\tensl}\rm}
\font\ninerm=cmr9 \font\sixrm=cmr6 \font\ninei=cmmi9 \font\sixi=cmmi6
\font\ninesy=cmsy9 \font\sixsy=cmsy6 \font\ninebf=cmbx9
\font\nineit=cmti9 \font\ninesl=cmsl9 \skewchar\ninei='177
\skewchar\sixi='177 \skewchar\ninesy='60 \skewchar\sixsy='60
\def\ninepoint{\def\rm{\fam0\ninerm}
\textfont0=\ninerm \scriptfont0=\sixrm \scriptscriptfont0=\fiverm
\textfont1=\ninei \scriptfont1=\sixi \scriptscriptfont1=\fivei
\textfont2=\ninesy \scriptfont2=\sixsy \scriptscriptfont2=\fivesy
\textfont\itfam=\ninei \def\it{\fam\itfam\nineit}\def\sl{\fam\slfam\ninesl}%
\textfont\bffam=\ninebf \def\bf{\fam\bffam\ninebf}\rm}
%
\hyphenation{anom-aly anom-alies coun-ter-term coun-ter-terms}

\global\newcount\subsubsecno \global\subsubsecno=0
\def\subsubsec#1\par{\global\advance\subsubsecno by1%
{\toks0{#1}\message{(\the\secno\the\subsecno\the\subsubsecno. \the\toks0)}}%
\ifnum\lastpenalty>9000\else\bigbreak\fi
\noindent{\it\hyperdef\hypernoname{subsubsection}{\the\secno.\the\subsecno\the\subsubsecno}%
{\the\secno.\the\subsecno.\the\subsubsecno.} #1}
\par\nobreak\medskip\nobreak\noindent\ignorespaces}

\def\DefWarn#1{}
\def\tikzcaption#1#2{\DefWarn#1\xdef#1{Fig.~\the\figno}
\writedef{#1\leftbracket Fig.\noexpand~\the\figno}%
{
\smallskip
\leftskip=20pt \rightskip=20pt \baselineskip12pt\noindent
{{\bf Fig.~\the\figno}\ \ninepoint #2}
\bigskip
\global\advance\figno by1 \par}}

\def\ntoalpha#1{%
\ifcase#1%
@%
\or A\or B\or C\or D\or E\or F\or G\or H\or I
\fi
}

\global\newcount\appno \global\appno=1
\def\applab#1{\xdef #1{\ntoalpha\appno}\writedef{#1\leftbracket#1}\wrlabeL{#1=#1}
\global\advance\appno by1}

\def\preprint#1 #2\par{\rightline{\vbox{\baselineskip12pt\hbox{#1}\hbox{#2}}}\vskip2cm}
%
\def\title#1\par{\centerline{\bf #1}\nopagenumbers\pageno=0}
\def\author#1\par{\bigskip\bigskip\centerline{#1}}

\newcount\addressno

\def\email#1#2{\unskip$^#1$\footnote{\null}{\kern-\parindent \llap{$^#1$\hskip1pt}email: #2}}

\def\startcenter{%
  \par
  \begingroup
  \leftskip=0pt plus 1fil
  \rightskip=\leftskip
  \parindent=0pt
  \parfillskip=0pt
}
\def\stopcenter{\endgroup}

\def\address{\bigskip%
  \ifnum\the\addressno=0\else\stopcenter\endgroup\fi
  \advance\addressno by 1%
  \begingroup
  \startcenter
  \it
  \obeylines
  \addressAux
}
\def\addressAux#1{#1}

\def\abstract{\stopcenter\endgroup\bigskip\bigskip\noindent}

\def\Dsl{\,\raise.15ex\hbox{/}\mkern-13.5mu D} 
\def\dsl{\raise.15ex\hbox{/}\kern-.57em\partial}
 
\def\boxeqn#1{\vcenter{\vbox{\hrule\hbox{\vrule\kern3pt\vbox{\kern3pt
	\hbox{${\displaystyle #1}$}\kern3pt}\kern3pt\vrule}\hrule}}}

\def\lie{\hbox{\it\$}} 

\def\a{\alpha}
\def\b{{\beta}}
\def\g{{\gamma}}
\def\d{{\delta}}

\def\l{\lambda}

\def\t{{\theta}}

\def\half{{1\over 2}}
\def\p{{\partial}}

\def\({\left(}
\def\){\right)}
\def\cF{{\cal F}}
\def\cW{{\cal W}}

\def\cA{{\cal A}}

\font\tenshuffle=shuffle10 \font\sevenshuffle=shuffle7 \font\fiveshuffle=shuffle7 at 5pt
\def\shuffle{{%
\def\Dshuffle{\mathbin{\hbox{\tenshuffle\char'001}}}%
\def\Sshuffle{\mathbin{\hbox{\sevenshuffle\char'001}}}%
\def\SSshuffle{\mathbin{\hbox{\fiveshuffle\char'001}}}%
\mathchoice{\Dshuffle}{\Dshuffle}{\Sshuffle}{\SSshuffle}}}


\def\qed{\hbox{\hskip 3pt
\vbox{\hrule\hbox to 7pt{\vrule height 7pt\hfill\vrule}
\hrule}}\hskip3pt}

\overfullrule=0pt\relax

\frenchspacing

\newread\instream \openin\instream= label.defs
\ifeof\instream \message{No labels in advance yet. Wait till next pass.}
\else \closein\instream \input label.defs
\fi
\writedefs

\def\arXiv:#1].{\hepthStrip#1 \nil}
\def\hepthStrip#1 #2\nil{\href{http://arxiv.org/abs/#1}{arXiv:#1 #2\unskip}].}

\input epsf
\input amssym

\font\frakfont=eufm10 at 10pt
\def\cK{{\cal K}}
\def\cG{{\cal G}}
\def\cH{{\cal H}}
\def\cX{{\cal X}}
\def\OOO{{\bf \Omega}}

\def\bW{{\Bbb W}}
\def\bF{{\Bbb F}}
\def\Box{\square}
\def\pplus{\phantom{+}}

\def\ce{\mathord{\hbox{\frakfont e}}}
\def\cf{\mathord{\hbox{\frakfont f}}}
\def\ch{\mathord{\hbox{\frakfont h}}}

\preprint DAMTP--2015--68

\vskip-30pt\relax
\title Non-linear gauge transformations in $D=10$ SYM theory and the BCJ duality

\author Seungjin Lee\email{\ddagger}{seungjin.lee@aei.mpg.de},
Carlos R. Mafra\email{\star}{mafra@ias.edu}$^\dagger$ and
	Oliver Schlotterer\email{\ddagger}{olivers@aei.mpg.de}

\address
$^\ddagger$Max--Planck--Institut f\"ur Gravitationsphysik
Albert--Einstein--Institut, 14476 Potsdam, Germany
\medskip
$^\star$Institute for Advanced Study, School of Natural Sciences,
Einstein Drive, Princeton, NJ 08540, USA
\medskip
$^\dagger$DAMTP, University of Cambridge
Wilberforce Road, Cambridge, CB3 0WA, UK

\abstract Recent progress on scattering amplitudes in super Yang--Mills and
superstring theory benefitted from the use of multiparticle superfields. They
universally capture tree-level subdiagrams, and their generating series solve the
non-linear equations of ten-dimensional super Yang--Mills.  We provide simplified
recursions for multiparticle superfields and relate them to earlier representations
through non-linear gauge transformations of their generating series. Moreover,
we discuss the gauge transformations which enforce their Lie symmetries as
suggested by the Bern--Carrasco--Johansson duality between color and
kinematics. Another gauge transformation due to Harnad and Shnider is shown to
streamline the theta-expansion of multiparticle superfields, bypassing the need to
use their recursion relations beyond the lowest components. The findings of this
work tremendously simplify the component extraction from kinematic factors in pure
spinor superspace.

\Date{October 2015}



\lref\BerendsME{
  F.A.~Berends and W.T.~Giele,
  ``Recursive Calculations for Processes with n Gluons,''
Nucl.\ Phys.\ B {\bf 306}, 759 (1988).
}
\lref\BerendsZN{
  F.A.~Berends and W.T.~Giele,
  ``Multiple Soft Gluon Radiation in Parton Processes,''
Nucl.\ Phys.\ B {\bf 313}, 595 (1989).
}
\lref\BerendsHF{
  F.A.~Berends, W.T.~Giele and H.~Kuijf,
  ``Exact and Approximate Expressions for Multi - Gluon Scattering,''
Nucl.\ Phys.\ B {\bf 333}, 120 (1990).
}
\lref\ree{
	R. Ree,
	``Lie elements and an algebra associated with shuffles'',
	Ann. Math. {\bf 68}, No. 2 (1958), 210--220.
}

\lref\otherpaper{
	C.R.~Mafra and O.~Schlotterer,
	``Berends--Giele recursions and the BCJ duality in superspace and components,''
	[arXiv:1510.08846 [hep-th]].
}

\lref\EOMBBs{
	C.R.~Mafra and O.~Schlotterer,
	``Multiparticle SYM equations of motion and pure spinor BRST blocks,''
	JHEP {\bf 1407}, 153 (2014).
	[arXiv:1404.4986 [hep-th]].
}

\lref\cohomology{
	C.R.~Mafra and O.~Schlotterer,
  	``Cohomology foundations of one-loop amplitudes in pure spinor superspace,''
	[arXiv:1408.3605 [hep-th]].
}

\lref\nptFT{
	C.R.~Mafra, O.~Schlotterer, S.~Stieberger and D.~Tsimpis,
	``A recursive method for SYM n-point tree amplitudes,''
	Phys.\ Rev.\ D {\bf 83}, 126012 (2011).
	[arXiv:1012.3981 [hep-th]].
}
\lref\nptTree{
	C.R.~Mafra, O.~Schlotterer and S.~Stieberger,
	``Complete N-Point Superstring Disk Amplitude I. Pure Spinor Computation,''
	Nucl.\ Phys.\ B {\bf 873}, 419 (2013).
	[arXiv:1106.2645 [hep-th]].
\semi
  	C.R.~Mafra, O.~Schlotterer and S.~Stieberger,
	``Complete N-Point Superstring Disk Amplitude II. Amplitude and Hypergeometric Function Structure,''
	Nucl.\ Phys.\ B {\bf 873}, 461 (2013).
	[arXiv:1106.2646 [hep-th]].
}

\lref\SelivanovHN{
  K.G.~Selivanov,
  ``On tree form-factors in (supersymmetric) Yang-Mills theory,''
Commun.\ Math.\ Phys.\  {\bf 208}, 671 (2000).
[hep-th/9809046].
}

\lref\wittentwistor{
	E.Witten,
        ``Twistor-Like Transform In Ten-Dimensions''
        Nucl.Phys. B {\bf 266}, 245~(1986).
}
\lref\psf{
 	N.~Berkovits,
	``Super-Poincare covariant quantization of the superstring,''
	JHEP {\bf 0004}, 018 (2000)
	[arXiv:hep-th/0001035].
}
\lref\BCJ{
	Z.~Bern, J.J.M.~Carrasco and H.~Johansson,
	``New Relations for Gauge-Theory Amplitudes,''
	Phys.\ Rev.\ D {\bf 78}, 085011 (2008).
	[arXiv:0805.3993 [hep-ph]].
}
\lref\KKref{
	R.~Kleiss and H.~Kuijf,
	``Multi - Gluon Cross-sections and Five Jet Production at Hadron Colliders,''
	Nucl.\ Phys.\ B {\bf 312}, 616 (1989)..
\semi
	V.~Del Duca, L.J.~Dixon and F.~Maltoni,
	``New color decompositions for gauge amplitudes at tree and loop level,''
	Nucl.\ Phys.\ B {\bf 571}, 51 (2000).
	[hep-ph/9910563].
}
\lref\towards{
	C.R.~Mafra,
	``Towards Field Theory Amplitudes From the Cohomology of Pure Spinor Superspace,''
	JHEP {\bf 1011}, 096 (2010).
	[arXiv:1007.3639 [hep-th]].
}
\lref\PolicastroVT{
  G.~Policastro and D.~Tsimpis,
  ``$R^4$, purified,''
Class.\ Quant.\ Grav.\  {\bf 23}, 4753 (2006).
[hep-th/0603165].
}

\lref\HarnadBC{
  J.P.~Harnad and S.~Shnider,
  ``Constraints And Field Equations For Ten-dimensional Superyang-mills Theory,''
Commun.\ Math.\ Phys.\  {\bf 106}, 183 (1986).
}

\lref\MafraGIA{
	C.R.~Mafra and O.~Schlotterer,
	``A solution to the non-linear equations of D=10 super Yang-Mills theory,''
	Phys.\ Rev.\ D {\bf 92}, no. 6, 066001 (2015).
	[arXiv:1501.05562 [hep-th]].
}

\lref\anomaly{
	N.~Berkovits and C.R.~Mafra,
	``Some Superstring Amplitude Computations with the Non-Minimal Pure Spinor Formalism,''
	JHEP {\bf 0611}, 079 (2006).
	[hep-th/0607187].
}

\lref\MafraKH{
  C.R.~Mafra and O.~Schlotterer,
  ``The Structure of n-Point One-Loop Open Superstring Amplitudes,''
JHEP {\bf 1408}, 099 (2014).
[arXiv:1203.6215 [hep-th]].
}

\lref\MafraKJ{
  C.~R.~Mafra, O.~Schlotterer and S.~Stieberger,
  ``Explicit BCJ Numerators from Pure Spinors,''
JHEP {\bf 1107}, 092 (2011).
[arXiv:1104.5224 [hep-th]].
}

\lref\MafraGJA{
  C.R.~Mafra and O.~Schlotterer,
  ``Towards one-loop SYM amplitudes from the pure spinor BRST cohomology,''
Fortsch.\ Phys.\  {\bf 63}, no. 2, 105 (2015).
[arXiv:1410.0668 [hep-th]].
}

\lref\MafraMJA{
  C.R.~Mafra and O.~Schlotterer,
  ``Two-loop five-point amplitudes of super Yang-Mills and supergravity in pure spinor superspace,''
  JHEP {\bf 1510}, 124 (2015).
[arXiv:1505.02746 [hep-th]].
}

\lref\MafraPN{
  C.~R.~Mafra,
  ``PSS: A FORM Program to Evaluate Pure Spinor Superspace Expressions,''
[arXiv:1007.4999 [hep-th]].
}

\lref\WWW{
	C.R.~Mafra and O.~Schlotterer,
	``PSS: From pure spinor superspace to components,''
	{\tt http://www.damtp.cam.ac.uk/user/crm66/SYM/pss.html}.
}

\lref\SiegelYI{
  W.~Siegel,
  ``Superfields in Higher Dimensional Space-time,''
Phys.\ Lett.\ B {\bf 80}, 220 (1979).
}

\lref\BernUE{
  Z.~Bern, J.~J.~M.~Carrasco and H.~Johansson,
  ``Perturbative Quantum Gravity as a Double Copy of Gauge Theory,''
Phys.\ Rev.\ Lett.\  {\bf 105}, 061602 (2010).
[arXiv:1004.0476 [hep-th]].
}

\lref\BernYG{
  Z.~Bern, T.~Dennen, Y.~t.~Huang and M.~Kiermaier,
  ``Gravity as the Square of Gauge Theory,''
Phys.\ Rev.\ D {\bf 82}, 065003 (2010).
[arXiv:1004.0693 [hep-th]].
}

\lref\BerkovitsRB{
  N.~Berkovits,
  ``Covariant quantization of the superparticle using pure spinors,''
JHEP {\bf 0109}, 016 (2001).
[hep-th/0105050].
}

\lref\BjornssonWM{
  J.~Bjornsson and M.~B.~Green,
  ``5 loops in 24/5 dimensions,''
JHEP {\bf 1008}, 132 (2010).
[arXiv:1004.2692 [hep-th]].
}

\lref\BjornssonWU{
  J.~Bjornsson,
  ``Multi-loop amplitudes in maximally supersymmetric pure spinor field theory,''
JHEP {\bf 1101}, 002 (2011).
[arXiv:1009.5906 [hep-th]].
}
\lref\ICTP{
  N.~Berkovits,
  ``ICTP lectures on covariant quantization of the superstring,''
[hep-th/0209059].
}
\lref\MPS{
        N.~Berkovits,
	``Multiloop amplitudes and vanishing theorems using the pure spinor formalism for the superstring,''
	JHEP {\bf 0409}, 047 (2004).
	[hep-th/0406055].
}
\lref\twoloop{
	N.~Berkovits,
  	``Super-Poincare covariant two-loop superstring amplitudes,''
	JHEP {\bf 0601}, 005 (2006).
	[hep-th/0503197].
\semi
	N.~Berkovits, C.R.~Mafra,
  	``Equivalence of two-loop superstring amplitudes in the pure spinor and RNS formalisms,''
	Phys.\ Rev.\ Lett.\  {\bf 96}, 011602 (2006).
	[hep-th/0509234].
}

\lref\GreenBZA{
  M.~B.~Green, C.R.~Mafra and O.~Schlotterer,
  ``Multiparticle one-loop amplitudes and S-duality in closed superstring theory,''
JHEP {\bf 1310}, 188 (2013).
[arXiv:1307.3534 [hep-th]].
}
\lref\ReutBook{
C.~Reutenauer,
``Free Lie Algebras'', London Mathematical Society Monographs, 1993.
}

\lref\CarrascoIWA{
  J.~J.~M.~Carrasco,
  ``Gauge and Gravity Amplitude Relations,''
[arXiv:1506.00974 [hep-th]].
}

\lref\Selivanov{
A.A.~Rosly and K.G.~Selivanov,
  ``On amplitudes in selfdual sector of Yang-Mills theory,''
Phys.\ Lett.\ B {\bf 399}, 135 (1997).
[hep-th/9611101].
\semi
	A.A.~Rosly and K.G.~Selivanov,
  	``Gravitational SD perturbiner,''
	[hep-th/9710196].
\semi
  K.G.~Selivanov,
  ``Postclassicism in tree amplitudes,''
[hep-th/9905128].
}
\lref\Bardeen{
  W.A.~Bardeen,
  ``Selfdual Yang-Mills theory, integrability and multiparton amplitudes,''
Prog.\ Theor.\ Phys.\ Suppl.\  {\bf 123}, 1 (1996).
}

\listtoc
\writetoc
\filbreak

\newsec Introduction

In recent years, the super-Poincar\'e covariant description \wittentwistor\ of
ten-dimensional super Yang--Mills theory (SYM) has been extensively used to compute
scattering amplitudes in string and field theory. This description features the
ten-dimensional superfields,
\eqn\nonlinearS{
\Bbb A_\a(x,\t),\;\; \Bbb A^m(x,\t),\;\; \Bbb W^\a(x,\t),\;\; \Bbb F^{mn}(x,\t)  \ ,
}
where $\Bbb A_\alpha,\Bbb A_m$ are the spinor and vector potentials
and $\Bbb W^\alpha, \Bbb F^{mn}$ their associated field-strengths.
They satisfy certain non-linear field equations to be reviewed below.

The appearance of the linearized versions $A_\a(x,\t)$, $A^m(x,\t)$, $W^\a(x,\t)$
and $F^{mn}(x,\t)$ of \nonlinearS\ in the massless vertex operators of the pure
spinor superstring \psf\ have brought these superfields to the forefront of
perturbation theory: They compactly encode the kinematic factors of scattering
amplitudes in string and field theory.

Following the standard CFT prescription for scattering amplitudes in the pure spinor superstring,
it soon became clear that the {\it linearized} superfields repeatedly appeared
in the same meaningful combinations. The study of short-distance singularities among massless
vertex operators gave rise to the notion of multiparticle superfields,
$$
K_P \in \{A_\a^P(x,\t),\; A^m_P(x,\t),\; W^\a_P(x,\t),\; F^{mn}_P(x,\t)\} \ .
$$
We gather the labels of several particles in $P=12 \ldots p$ and collectively refer to the four types of
superfields via $K_P$ to avoid the cluttering of Lorentz indices.

In the last years, two distinct ways of obtaining the explicit expressions
of multiparticle superfields have been proposed. In 2011 and 2012
\refs{\nptTree,\MafraKH}, their construction
closely followed the (lengthy) OPE calculations in superstring amplitudes, 
leading to expressions for $K_P$ which
satisfy the Lie symmetries of nested commutators $[\ldots [[t^1,t^2],t^3],\ldots ,t^ p]$
under permutations of the labels in $P=12\ldots p$. In 2014 \EOMBBs, an
efficient recursive definition of multiparticle superfields was given in terms of a
cubic-vertex prescription $K_{[P,Q]}$, bypassing the need to perform OPEs beyond multiplicity $p=2$. A chain of redefinitions
was supplemented in order to recover the same Lie symmetries as in the previous approach.

In addition to the (local) multiparticle superfields, the superstring
amplitude calculations also suggested natural definitions of their
non-local counterparts, called {\it Berends--Giele currents} and
represented by calligraphic letters,
\eqn\nonumber{
\cK_P \in \{\cA^P_\a(x,\t),\;\cA^P_m(x,\t),\; \cW_P^\a(x,\t),\; \cF_P^{mn}(x,\t)\} \ .
}
As described in \refs{\nptTree,\EOMBBs}, the precise definition of $\cK_P$ used an intuitive
mapping between planar binary trees (or cubic graphs) and 
Lie symmetry-satisfying multiparticle superfields, dressed with the propagators of the graph.
These Berends--Giele currents elegantly capture kinematic
factors of multiparticle amplitudes in both string and field-theory.

As one of the main result of this article, we provide an alternative definition of
Berends--Giele currents which tremendously simplifies the construction of
earlier work \EOMBBs\ while preserving their equations of motion.

\subsec Generating series and non-linear gauge transformations

A new perspective on multiparticle superfields $K_P$ and their associated Berends--Giele currents ${\cal K}_P$
is provided by the generating series of Berends--Giele currents. These generating series are
an expansion in terms of Lie-algebra generators $t^i$ 
with multiparticle Berends--Giele currents as coefficients \MafraGIA,
\eqn\introGenSer{
\Bbb K \equiv \sum_{p=1}^{\infty} \sum_{i_1,i_2,\dots,i_p}  \cK_{i_1 i_2 \dots i_p}
t^{i_1} t^{i_2} \ldots t^{i_p} \ .
}
As a key feature of these generating series $\Bbb K \in \{\Bbb A_\a(x,\t),\Bbb A^m(x,\t), \Bbb W^\a(x,\t),\Bbb
F^{mn}(x,\t)\}$, they are Lie algebra-valued and solve the non-linear field
equations of ten-dimensional SYM theory. 
These equations are invariant under non-linear gauge transformations \wittentwistor,
\eqnn\introGauge
$$\eqalignno{
\delta_{\OOO } \Bbb A_\alpha &
   = \big[D_\alpha ,\OOO \big] - \big[ \Bbb A_\alpha ,\OOO \big]  \,, \quad
 \, \; \delta_{\OOO }   \Bbb W^\alpha =  \big[\OOO, \Bbb W^\alpha  \big]\,,
   &\introGauge
\cr
\delta_{\OOO }  \Bbb A_m  
 &= \big[\partial_m ,\OOO \big] - \big[\Bbb A_m ,\OOO \big]
\,, \quad
 \delta_{\OOO }  \Bbb F^{mn} =  \big[\OOO, \Bbb F^{mn}  \big]   \,,
}
$$
where $\OOO(x,\t)$ is a generating series of multiparticle gauge parameters $\Omega_P$.
This non-linear gauge invariance will be the main topic of this work.
It underpins the earlier constructions
of multiparticle superfields and provides a surprising
link between the Bern--Carrasco--Johansson (BCJ)
duality \refs{\BCJ,\BernYG,\BernUE} and multiparticle gauge transformations.

\subsec Non-linear gauge transformations and the BCJ duality

As will be shown in this paper,
the cubic-vertex prescription $K_{[P,Q]}$ appearing in the earlier
construction of multiparticle superfields \EOMBBs\
turns out to have a direct
non-local counterpart for Berends--Giele currents
\eqn\BGLorentz{
\cK_P \equiv {1\over s_P}\sum_{XY=P}\cK_{[X,Y]}
}
with the same functional form for the currents $\cK_{[X,Y]}$ as seen for the local
fields $K_{[X,Y]}$. The recursive definition \BGLorentz\ yields a particular gauge
where $k^P_m\cA^m_P(x,\t) = 0$, in other words, the generating series $\Bbb
A_m^{\rm L}$ of the currents in \BGLorentz\ realizes {\it Lorentz gauge}.

The redefinitions required by imposing the Lie symmetries on the multiparticle
superfields in the previous constructions \refs{\nptTree,\EOMBBs}
are now understood as a change of gauge. Starting from the
definitions in the Lorentz gauge as above, the superfield
redefinitions discussed in \refs{\nptTree,\EOMBBs} amount to enforcing
the {\it BCJ gauge}, e.g.,
\eqn\relAm{
\Bbb A_m^{\rm BCJ} =
\Bbb A^{\rm L}_m + \big[\partial_m ,\OOO^{\rm BCJ} \big] - \big[\Bbb A^{\rm L}_m ,\OOO^{\rm BCJ} \big] \ ,
}
where the superscripts ${}^{\rm BCJ}$ and ${}^{\rm L}$ refer to the redefined
superfields of \refs{\nptTree,\EOMBBs} and the new recursive constructions
discussed in this paper. The gauge parameter\foot{For historical reasons, $\OOO^{\rm BCJ}$ 
will be denoted by $-\Bbb H$ in section 3.} $\OOO^{\rm BCJ}$ in the sense of \introGauge\
will be described in section 3, with complete expressions up to the fifth order in
the multiparticle expansion.

The terminology ``BCJ gauge'' for the above transformations is motivated
by the BCJ conjecture \BCJ\
 on a duality between color and kinematics:
The kinematic factors $N_i$ of scattering amplitudes can be arranged to
satisfy the same Jacobi identity as their associated color factors $C_i$, see \BernYG\ for the striking
impact on gravity amplitudes, \BernUE\ for the loop-level formulation of the conjecture 
and \CarrascoIWA\ for a review. Incidentally, the family $K_P^{\rm BCJ}$ of
multiparticle superfields in the BCJ gauge satisfies
the same ``generalized Lie symmetries'' \ReutBook\ as a string of structure constants in $[t^a ,t^b ]= f^{abc} t^c$,
\eqn\LieIntro{
{\rm ``kinematics"} \ K^{\rm BCJ}_{12\ldots p}\ \longleftrightarrow f^{1 2 a_3} f^{a_3 3 a_4} f^{a_4 4
a_5} \ldots f^{a_p p a_{p+1}} \ {\rm ``color"} \, .
}
The relation between the tree-level BCJ duality
and the superfields in the BCJ gauge
can be seen from the tree-level
amplitudes computed with the pure spinor superstring \MafraKJ.

At tree level, the numerators $N_i$ are assembled from cubic expressions $A^P_\a A^Q_\beta A^R_\gamma$
where the particular linear combinations of multiparticle labels $P,Q,R$ follow
from the field-theory limit of the superstring amplitude, see \figBCJtree. As shown in \MafraKJ, the numerators
resulting from this procedure obey the color-kinematics duality for any number of
external particles. The superfields in the ``BCJ gauge'' were
an essential requirement in the derivation of BCJ-satisfying numerators
from the pure spinor superstring\foot{In eliminating spurious double poles from the string computation, BCJ 
gauge of the multiparticle superfields is automatically attained \refs{\nptTree}.}. Non-linear 
gauge transformations of the generating series \introGenSer\ of multiparticle
superfields reparametrize the SYM amplitudes by moving terms between different cubic diagrams.
They can therefore be viewed as an example of the ``generalized gauge freedom'' of \refs{\BCJ, \BernYG, \BernUE}.

\ifig\figBCJtree{The basis of half-ladder diagrams with legs 1 and $n-1$ attached to opposite
endpoints furnish the manifestly-local pure spinor representation of tree-level
numerators $V_{12 \ldots j}V_n V_{n-1,n-2, \ldots j+1}$ built from
SYM superfields in the BCJ gauge.}
{\epsfxsize=.62\hsize\epsfbox{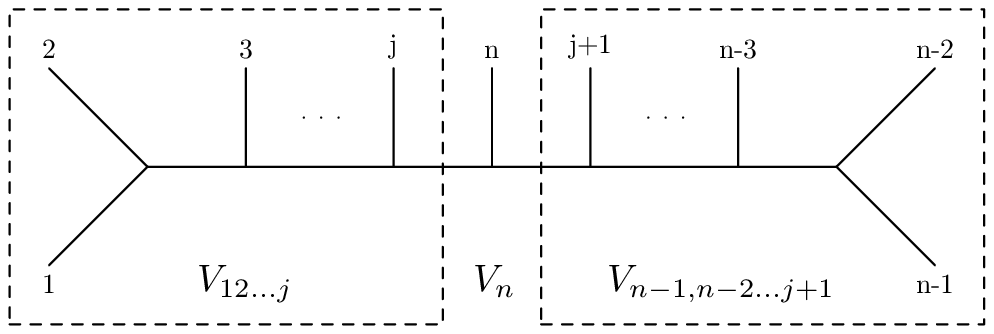}}

\ifig\figMONSTER{
The pure spinor expressions of arbitrary box and double-box numerators are given
by certain multiparticle building blocks $V_A T_{B,C,D}$ \MafraGJA\ and $T_{A,B|C,D}$ \MafraMJA. They
furnish a manifestly local representation that satisfies the BCJ identities within
each external tree subdiagram when the SYM superfields are in the BCJ gauge.
}
{\epsfxsize=.60\hsize\epsfbox{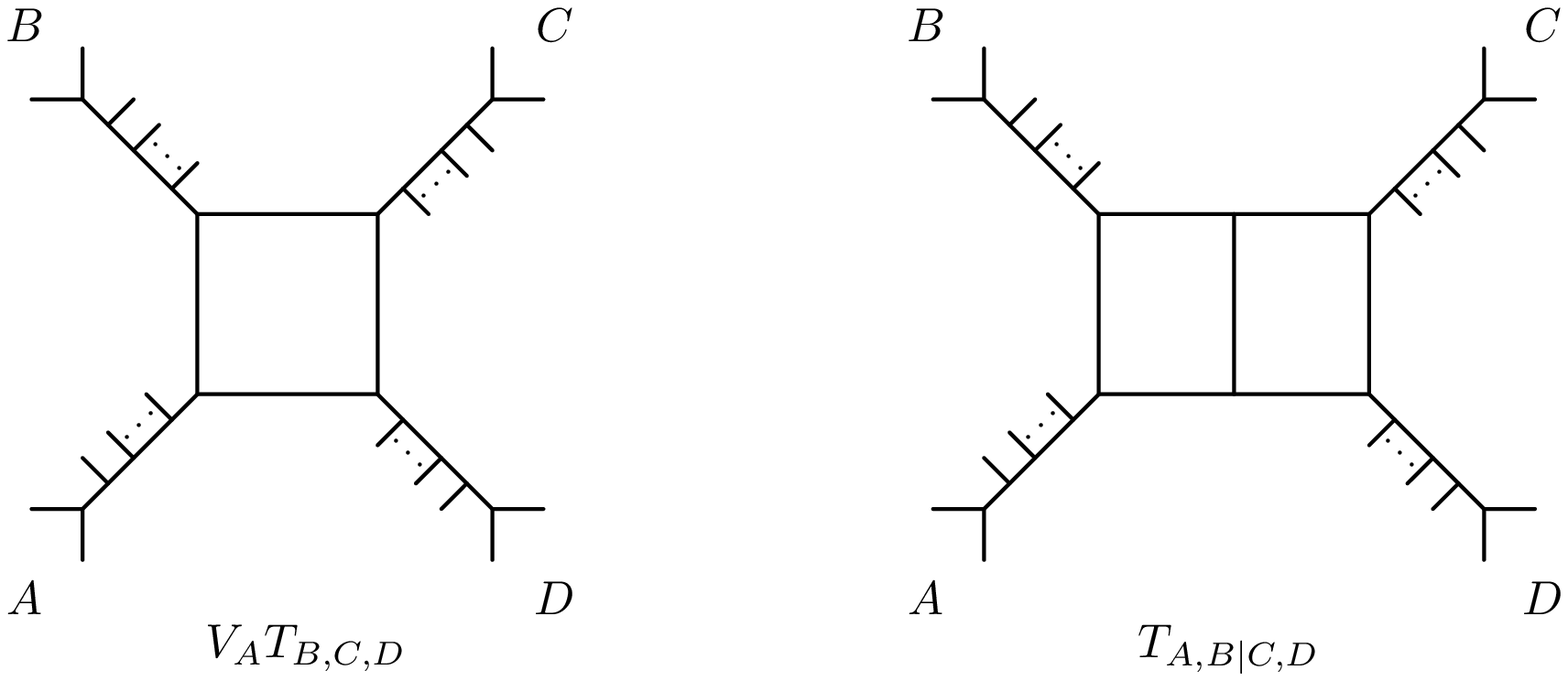}}

At loop level, BCJ-satisfying five-point integrands at both one-
and two-loops were recently derived using multiparticle superfields in the
BCJ gauge \refs{\MafraGJA, \MafraMJA}\foot{It should be pointed out
that the straightforward derivation of the six-point integrand
at one-loop does not satisfy the BCJ duality \MafraGJA. Although not conclusive,
the failure seems to be related to the well-known six-point gauge anomaly and
deserves further investigation.}. At any multiloop order, kinematic Jacobi identities within
tree-level subdiagram are manifestly satisfied if they are represented by multiparticle
superfields in BCJ gauge. This for instance applies to the general box and double-box 
diagram displayed in \figMONSTER\ where the multiparticle labels $A,B,C$ and $D$
refer to appropriate superfields with the symmetry \LieIntro. The ubiquitous
appearance of multiparticle superfields calls for an efficient handle on their components,
i.e.~their dependence on the Grassmann-odd superspace coordinates $\theta^\alpha$. 

\subsec Theta-expansions in Harnad--Shnider gauge

\noindent In the same way as the Lie symmetries required by the BCJ duality could
be attained through a non-linear gauge transformation \relAm, we will simplify the
theta-expansion of Berends--Giele currents through a convenient choice of
multiparticle gauge parameters. The underlying gauge condition $\theta^\alpha \Bbb
A_\alpha^{\rm HS}=0$ goes back to Harnad and Shnider (HS) \HarnadBC\ and has been
further studied in the context of linearized superfields \PolicastroVT. We apply
this line of thoughts to the multiparticle level and obtain economic
theta-expansions for Berends--Giele currents ${\cal K}_P$ which largely resemble
the linearized counterparts. Non-linear deviations at higher powers of theta are
controlled by superfields of higher mass dimension \MafraGIA.

The theta-expansions in HS gauge significantly alleviate the conversion of
kinematic factors in pure spinor superspace to their components involving gluons
and gluinos. The computational effort caused by large numbers of external states
\MafraPN\ can be tremendously reduced, and the resulting structural insights into
the tree-level components are discussed in a companion paper \otherpaper. A huge
long-term benefit for higher orders in perturbation theory is expected from the 
quick access to the component information on multiloop kinematic factors.

\subsec Outline

\noindent This paper is organized as follows. In section~2, the field equations of
ten-dimensional SYM are reviewed and exploited to construct Berends--Giele currents
in Lorentz gauge. Their gauge equivalence to the earlier construction of \EOMBBs\
in BCJ gauge is clarified in section~3. In section~4, the key ideas of HS gauge are
reviewed and applied to streamline the theta-expansions of Berends--Giele currents,
starting from either Lorentz gauge or BCJ gauge. Finally, we conclude in section 5
with applications of the improved theta-expansions to scattering amplitudes in pure
spinor superspace.

\newsec Super-Poincare description of ten-dimensional super Yang--Mills

\subsec Non-linear super Yang--Mills

Ten-dimensional SYM can be described in a super-Poincar\'e covariant manner using
superspace coordinates $\{ x^m,\t^\a\}$ where $m,n=0, \ldots,9$ and $\a,\b=1, \ldots,
16$ denote vector and spinor indices of the Lorentz group.
Using Lie algebra-valued connections $\Bbb
A_\a = \Bbb A_\alpha(x,\t)$ and $\Bbb A_m = \Bbb A_m(x,\t)$, one defines
supercovariant derivatives \refs{\SiegelYI, \wittentwistor},
\eqn\covder{
\nabla_\alpha \equiv D_\alpha - \Bbb A_\alpha\,, \ \
\ \ \nabla_m \equiv \p_m - \Bbb A_m \ .
}
The fermionic differential operators
\eqn\deralg{ D_\a \equiv {\partial \over \partial \theta^\a}
+ {1\over 2}(\gamma^m\theta)_\a\p_m\,, \qquad \{D_\a,D_\b\} = \g^m_{\a\b}\partial_m 
}
involve the $16\times 16$ Pauli matrices $\g^m_{\a\b}=\g^m_{\b\a}$ subject to the
Clifford algebra $\g^{(m}_{\a\b}\g^{n)\b\g} = 2\eta^{mn}\d_\a^\g$, and the
convention for (anti)symmetrizing indices does not include~${1\over 2}$.
The constraint equation
$\big\{ \nabla_\alpha ,\nabla_\beta \big\} = \gamma^m_{\alpha \beta} \nabla_m$
together with Bianchi identities then
lead to the non-linear equations of motion \wittentwistor,
\eqn\SYMeomO{
\eqalign{
 \big\{ \nabla_{\alpha} , \nabla_{\beta} \big\} &= \gamma^m_{\alpha \beta}
 \nabla_m\,, \cr
\big[ \nabla_\alpha , \nabla_m\big] &= - (\gamma_m \bW)_\alpha\,,
}
\qquad\eqalign{
\big\{ \nabla_{\alpha} ,\bW^\beta \big\} &= {1\over 4}
(\gamma^{mn})_{\alpha}{}^{\beta} \bF_{mn}\,, \cr
\big[\nabla_{\alpha}, \bF^{mn} \big] &=  (\Bbb W^{[m} \gamma^{n]})_\alpha \,,
}}
where
\eqn\WWWdef{
\Bbb F_{mn} \equiv - \big[  \nabla_m, \nabla_n \big] \ , \qquad
\Bbb W^{\alpha}_m \equiv \big[  \nabla_m, \Bbb W^\alpha \big]
\,.
}
Equivalently, using the definitions \covder\ the equations of motion \SYMeomO\ become
\eqn\SYMeom{
\eqalign{
\{D_{(\alpha},\Bbb A_{\beta) } \}&=\phantom{{1\over4}}\gamma^{m}_{\alpha\beta}\Bbb A_m+\{\Bbb A_\alpha, \Bbb A_\beta\} \ ,\cr
[D_{\alpha},\Bbb A_m]&=[\partial_m,\Bbb A_\alpha]+(\gamma_m \Bbb W)_\alpha+[\Bbb A_\alpha,\Bbb A_m] \ ,
}\quad\eqalign{
\{D_\alpha,\Bbb W^\beta\}&={1\over4}(\gamma^{mn})_\alpha^{\phantom{\alpha}\beta}\Bbb F_{mn}+\{\Bbb A_\alpha,\Bbb W^\beta\} \cr
[D_\alpha, \Bbb F^{mn}]&= (\Bbb W^{[m} \gamma^{n]})_\alpha +[\Bbb A_\alpha,\Bbb
F^{mn}]\,.
}}
It is straightforward to check that \SYMeomO\ or \SYMeom\ are preserved
by the non-linear gauge transformations,
\eqnn\NLgauge
$$\eqalignno{
\delta_{\OOO } \Bbb A_\alpha &
   = \big[\nabla_\alpha ,\OOO \big] , \quad \ \,  \delta_{\OOO }  \Bbb A_m  
 = \big[\nabla_m ,\OOO \big]
   &\NLgauge
\cr
 \delta_{\OOO }   \Bbb W^\alpha &=  \big[\OOO, \Bbb W^\alpha  \big] 
 ,\quad \delta_{\OOO }  \Bbb F^{mn} =  \big[\OOO, \Bbb F^{mn}  \big]
  ,\quad \delta_{\OOO }  \Bbb W^{m\alpha} =  \big[\OOO, \Bbb W^{m\alpha}  \big]   \ ,
}
$$
with Lie algebra-valued gauge parameter $\OOO=\OOO(x,\theta)$.

\subsubsec Linearized super Yang--Mills

\noindent Discarding the quadratic terms in the superfields from the
equations of motion \SYMeom\ yields
the field equations of linearized SYM,
\eqn\SYM{
\eqalign{
\{ D_{(\a} , A_{\b)} \}  &= \phantom{{1\over 4}}\g^m_{\a\b} A_m,\cr
[ D_\a, A_m ] &= (\g_m W)_\a + [\partial_m ,A_\a] ,\cr
}
\qquad\eqalign{
\{D_\a ,W^{\b} \} &= {1\over 4}(\g^{mn})_\a{}^\b F_{mn}\cr
 [ D_\a , F_{mn} ] &= [\partial_{[m}, (\g_{n]} W)_\a]\ .
}}
They are invariant under the linearized gauge transformations,
\eqn\lingauge{
\delta_{\Omega} A_\alpha = \big[D_\alpha ,\Omega \big] \,,
\quad \ \delta_{\Omega}  A_m = \big[\partial_m ,\Omega \big] \,,
\quad  \ \delta_{\Omega}     W^\alpha =  0  \,,\quad \ \delta_{\Omega}  F^{mn} = 0 \ .
}
Note that the massless vertex operators in the open pure spinor superstring \psf\ are given
in terms of these linearized superfields, and the equations of motion \SYM\ imply
their BRST invariance \ICTP.

\subsec Supersymmetric Berends--Giele currents in Lorentz gauge
\par\subseclab\simprec

\noindent
For a multiparticle label
$P\equiv i_1 i_2 i_3\ldots i_p$ with each $i_j$ referring to an external SYM state,
we define a set of multiparticle {\it Berends--Giele currents\/} 
\eqn\calKdef{
\cK_P \in \{\cA^P_\a,
\cA^m_P, \cW^\a_P,\cF^{mn}_P\}
}
as follows. The single-particle currents $\cK_i$ are given by the linearized
superfields,
$\cK_i \in \{A^i_\a, A^m_i, W^\a_i, F^{mn}_i\}$,
while multiparticle instances follow from the recursion\foot{This definition
of the supersymmetric Berends--Giele currents
closely generalizes the standard Berends--Giele currents $J^m_P$ of \BerendsME.
When the fermions are set to zero, $J^m_P$
can be identified as the theta-independent component of $\cA^m_P(x,\t)$. Furthermore,
the quartic-vertex interaction $\{J_X,J_Y,J_Z\}$ of \BerendsME\ is
automatically included in the cubic-vertex prescription $\cK_{[X,Y]}$ \otherpaper.}
\eqn\BGdef{
\cK_P \equiv {1\over s_{P}}\sum_{XY=P}\cK_{[X,Y]} \,, 
}
where
\eqnn\cAalpha
\eqnn\cAm
\eqnn\cWalpha
\eqnn\cFmn
$$\eqalignno{
\cA^{[P,Q]}_\a &\equiv - \half\bigl[  \cA^{P}_\a (k^{P}\cdot  \cA^Q)
+  \cA^{P}_m (\g^m \cW^Q)_\a - (P\leftrightarrow Q)\bigr]  &\cAalpha\cr
\cA^{[P,Q]}_m &\equiv - \half\bigl[ \cA^{P}_m (k^P\cdot  \cA^{Q}) + \cA^{P}_n  \cF^Q_{mn}
- ( \cW^{P}\g_m \cW^Q)
- (P \leftrightarrow Q)\bigr] &\cAm\cr
\cW_{[P,Q]}^\a &\equiv  \half (k_P^m + k_Q^m) \gamma_m^{\alpha \beta}
 \big[ \cA_P^n (\gamma_n \cW_Q)_\beta  - (P \leftrightarrow Q) \big]
&\cWalpha\cr
\cF^{mn}_P &\equiv k_P^m \cA_P^n - k_P^n \cA_P^m
- \sum_{XY=P}\!\!\big( \cA_X^m \cA_Y^n - \cA_X^n \cA_Y^m \big) \ .
&\cFmn\cr 
}$$
Multiparticle momenta as well as their associated Mandelstam invariants are defined by
\eqn\multmand{
k_P^m \equiv k^m_{i_1}+k^m_{i_2}+\cdots + k^m_{i_p}
\ , \ \ \ \ \ \
s_P\equiv{1\over 2}k_P^2 \ ,
}
and the sum over multiparticle labels $XY=P$ in \BGdef\ and \cFmn\ instructs to deconcatenate $P=i_1 i_2
i_3\ldots i_p$ into non-empty words $X=i_1 i_2\ldots i_j$ and $Y= i_{j+1}\ldots
i_p$ with $j=1,2,\ldots,p-1$. Alternative recursive formul{\ae} for $\cW_P^{\alpha}$ and
$\cF^{mn}_P$ read\foot{The recursion for Berends--Giele currents ${\cal W}^\alpha_P$ and
${\cal F}^{mn}_P$ based on \altW\ is actually closer to the original string-inspired
construction of multiparticle superfields where the key input stems from the short-distance
behaviour of integrated vertex operators \EOMBBs.}
\eqnn\altW
$$\eqalignno{
\cW_{[P,Q]}^\a &= - \half \bigl[  \cW_{P}^\a (k_P\cdot \cA_Q) + \cW_P^{m\a} \cA^m_Q
+ {1\over 2}(\g_{rs} \cW_P)^\a  \cF_{Q}^{rs}  
- (P \leftrightarrow Q) \bigr]&\altW\cr
\cF^{mn}_{[P,Q]} &= - {1\over 2} \big[ {\cal F}^{mn}_P (k_P \cdot {\cal A}_Q)
+ {\cal F}_{P}^{p|mn} {\cal A}_p^Q + 2 {\cal F}_P^{mp} {\cal F}_{Q \, p}^n
+ 4 \g^{[m}_{\alpha \beta} {\cal W}_P^{n]\alpha} {\cal W}_Q^\beta
- (P \leftrightarrow Q) \big] \ ,
}
$$
with superfields $\cW_P^{m \alpha}, \cF_P^{m |pq}$ of higher mass dimension,
\eqnn\cWm
\eqnn\cFmpq
$$\eqalignno{
\cW_P^{m \alpha} &\equiv  k_P^{m} {\cal W}_{P}^{\alpha}+ \sum_{XY=P}\!\! \bigl(  {\cal W}_{X}^{\alpha} {\cal A}^{m}_Y
-{\cal W}_{Y}^{\alpha} {\cal A}^{m}_X\bigr)  &\cWm \cr
\cF_P^{m |pq} &\equiv  k_P^{m} {\cal F}_{P}^{pq}+ \sum_{XY=P}\!\! \bigl(  {\cal F}_{X}^{pq} {\cal A}^{m}_Y
-{\cal F}_{Y}^{pq} {\cal A}^{m}_X\bigr)  \ .
&\cFmpq 
}$$
One can show by induction that the Berends--Giele currents defined
in \cAalpha\ to \cFmn\ obey the equations of motion
\eqnn\BGEOM
$$\eqalignno{
D_{(\alpha} \cA_{\beta)}^P &=
\g^m_{\alpha \beta} \cA_m^P
+ \! \! \sum_{XY=P} \! \! \bigl( \cA_\alpha^X \cA_{\beta}^Y
-\cA_\alpha^Y \cA_{\beta}^X\bigr) &\BGEOM\cr
D_\alpha \cA_m^P &= k^P_m \cA_{\alpha}^P + (\g_m \cW_P)_\alpha + \!\! \!
\sum_{XY=P} \! \! \bigl( \cA_\alpha^X \cA_{m}^Y -\cA_\alpha^Y \cA_{m}^X\bigr)
\cr
D_{\alpha} \cW^\beta_P &= {1\over 4} (\g^{mn})_{\alpha}{}^{\beta} {\cal F}^P_{mn}
+ \! \! \sum_{XY=P}  \! \!\bigl( \cA_\alpha^X \cW_Y^{\beta} -\cA_\alpha^Y \cW_X^\beta\bigr)
\cr
D_{\alpha} {\cal F}^{mn}_P  &= (\cW^{[m}_P \g^{n]} )_\alpha 
+\!\! \sum_{XY=P} \!\!\bigl( \cA_\a^X {\cal F}_Y^{mn} -\cA_\alpha^Y {\cal F}_X^{mn}\bigr)  \ .
}$$
Apart from the terms along with the deconcatenation sum $\sum_{XY=P}$, these
multiparticle equations of motion have the same form as the linearized ones \SYM. 
They play a key role for the BRST invariance  of scattering  amplitudes
in string  and field theory, see \refs{\towards, \nptFT, \nptTree}  for examples  at
tree-level and \refs{\MafraKH, \GreenBZA,  \MafraGJA, \MafraMJA} at loop-level. The
need for such objects was also observed in the
worldline version of the pure spinor formalism \refs{\BjornssonWU, \BjornssonWM}.

In addition, one can also show by induction that the currents defined in \BGdef\
satisfy,
\eqnn\tmpt
\eqnn\tmpr
\eqnn\tmps
$$\eqalignno{
k^P_m \cA_P^m &= 0 &\tmpt\cr
k_m^P (\g^m \cW_P)_\alpha & =\!\!\sum_{XY=P} \! \! \big[ \cA_m^X (\g^m \cW_Y)_\a
- \cA_m^Y (\g^m \cW_X)_\a\big] &\tmpr\cr
k_m^P \cF^{mn}_P &= \!\!\sum_{XY=P} \!\!\bigl[  2(\cW_X \g^n \cW_Y)
+ \cA_m^X \cF^{mn}_Y- \cA_m^Y \cF^{mn}_X\bigr]\,. &\tmps\cr
}$$
As we will see, \tmpt\ implies that the generating series of Berends--Giele currents \BGdef\ is in {\it Lorentz gauge}.

\subsubsec Symmetries of supersymmetric Berends--Giele currents

The currents $\cK_P(x,\t)$ defined above satisfy the following symmetry proven in appendix~\proofshuffle,
\eqn\shuff{
{\cal K}_{A \shuffle B}= 0,  \quad \forall A,B \neq \emptyset\ ,
}
where the shuffle product $\shuffle$ between the words $A=a_1 a_2
\ldots a_{|A|}$ and $B= b_1 b_2\ldots b_{|B|}$ is defined recursively by
\eqn\Shrecurs{
\emptyset\shuffle A = A\shuffle\emptyset = A,\qquad
A\shuffle B \equiv a_1(a_2 \ldots a_{|A|} \shuffle B) + b_1(b_2 \ldots b_{|B|}
\shuffle A)\,,
}
and $\emptyset$ denotes the empty word.

As elaborated in a companion paper \otherpaper, setting the fermions
to zero reduces the theta-independent component of $\cA_P^m(x,\t)$ to the gluonic
current $J^m_P$ defined by Berends and Giele \BerendsME, thus \shuff\ implies
the symmetry $J^m_{A\shuffle B}=0$ derived in \BerendsZN. These facts explain why $\cK_P(x,\t)$ are
called supersymmetric Berends--Giele currents.

\subsec Generating series of Berends--Giele currents
\par\subseclab\lorentzsec

\noindent The generating series of multiparticle Berends--Giele currents ${\cal K}_P \in \{ {\cal A}_\alpha^P, {\cal A}^m_P, {\cal W}_P^\alpha,\ldots\}$
\eqn\LieEldef{
\Bbb K \in\{\Bbb A_\alpha,\Bbb A^m, \Bbb W^\alpha, \ldots \}
}
is an expansion in terms of Lie-algebra generators $t^{i_j}$ \MafraGIA
\eqnn\series{
$$\eqalignno{
\Bbb K & \equiv \sum_{p=1}^{\infty} \sum_{i_1,i_2,\dots,i_p}  \cK_{i_1 i_2 \dots i_p} t^{i_1} t^{i_2} \ldots t^{i_p} &\series
\cr
&= \sum_{p=1}^{\infty} \sum_{i_1,i_2,\dots,i_p} {1\over p}\ \cK_{i_1 i_2 \dots i_p} [t^{i_1},[t^{i_2},\dots,[t^{i_{p-1}},t^{i_p}]]\dots ]\ .
}$$
The second line follows from the Berends--Giele symmetry \shuff\ and guarantees that
$\Bbb K$ is Lie algebra valued, see \ree\ for a proof.
The equations of motion \BGEOM\ satisfied by the Berends--Giele
currents imply that $\Bbb K$
satisfies the non-linear field equations \SYMeom\ \MafraGIA\foot{The notion of a
generating series which
solves the field equations
and gives rise to tree amplitudes is also central to
the ``perturbiner'' formalism \Selivanov. This approach has been
used to derive a generating series of Yang--Mills MHV amplitudes, see \SelivanovHN\ for a supersymmetric extension. 
However, the generic Yang--Mills amplitudes have never been obtained (see also \Bardeen).
We thank Nima Arkani-Hamed for pointing out these references.
}.

Given that the Mandelstam invariant $s_P$ in \multmand\ arises from half the d'Alembertian,
\eqn\boxop{
\Box \Bbb K  \equiv \big[ \partial^m , [\partial_m , \Bbb K ] \big] \ ,
}
the recursive prescriptions \cAalpha\ to \cWalpha\ for ${\cal A}_\alpha^P,{\cal A}_m^P,{\cal W}^\alpha_P$
can be reexpressed at the level of the generating series as
\eqnn\boxAalpha
\eqnn\boxAm
\eqnn\boxWalpha
$$\eqalignno{
\Box \Bbb A_{\alpha} &=
\big[ \Bbb A_m ,[\partial^m ,\Bbb A_\alpha]\big] + \big[ (\gamma^m \Bbb W)_\alpha, \Bbb A_m \big]    &\boxAalpha
\cr
\Box \Bbb A_m &=
 \big[ \Bbb A_p , [\partial^p, \Bbb A^m]  \big]  + \big[ \Bbb F^{mp}, \Bbb A_p \big]
 +\gamma^m_{\alpha \beta} \{ \Bbb W^\alpha, \Bbb W^\beta \}  &\boxAm
\cr
\Box \Bbb W^{\alpha} &= \big[ \partial_m , [ \Bbb A_n ,  (\gamma^m \gamma^n \Bbb  W)^\alpha]] \ .
&\boxWalpha
}$$
As detailed in the following subsection, these second-order differential equations can be verified from
\SYMeom\ and \WWWdef, provided that Lorentz gauge is imposed,
\eqn\Lorentz{
[ \partial_m , \Bbb A^m ] = 0 \ .
}
Similar manipulations lead to the generating-series representation of \altW,
\eqnn\altboxW
\eqnn\altboxF
$$\eqalignno{
\Box \Bbb W^{\alpha} &=\big[ \Bbb A_m, [ \partial^m, \Bbb W^\alpha ]  \big] + \big[ \Bbb A^m, \Bbb W^\alpha_m  \big]
+ {1\over 2} \big[\Bbb F_{mn}, (\gamma^{mn} \Bbb W)^\alpha \big]  &\altboxW \cr
\Box \Bbb F^{mn} &= [\Bbb A_p, [\partial^p ,\Bbb F^{mn}]] + [\Bbb A_p,  \Bbb F^{p|mn}]
+ 2 [\Bbb F^{mp}, \Bbb F_p{}^n ] + 4 \{ (\Bbb W^{[m} \g^{n]})_\alpha , \Bbb
W^\alpha \}\,,
&\altboxF
}$$
where $\Bbb F^{p|mn}$ denotes the generating series of \cFmpq.
Equivalence of \altboxW\ and \boxWalpha\ follows from the Dirac equation,
\eqn\DiracW{
\nabla_m (\gamma^m \Bbb W)_\alpha=0\,,
}
i.e. the generating series of \tmpr. In summary, the recursive prescriptions
\BGdef\ to \cFmn\ for multiparticle
superfields yield a solution of the SYM equations in Lorentz gauge.

\subsubsec Deriving non-linear wave equations

We shall now derive the non-linear wave equations \boxAalpha, \boxAm, \altboxW\ and \altboxF\
for the non-linear superfields $\Bbb K$ in Lorentz gauge. By Jacobi identities and
repeated use of $\partial^m = \nabla^m + \Bbb A^m$, we have
\eqnn\boxone
$$\eqalignno{
\Box \Bbb K &= [\nabla^m + \Bbb A^m , [\partial_m,\Bbb K ]] &\boxone \cr
&= [ [ \nabla^m , \partial_m ] , \Bbb K] + [ \Bbb A^m , [\partial_m,\Bbb K ]]
+ [ \Bbb A^m , [\nabla_m,\Bbb K ]] + [ \nabla^m , [\nabla_m,\Bbb K ]] \ .
}$$
The first term in the second line vanishes in Lorentz gauge \Lorentz\ by
$[  \partial_m , \nabla^m ] =-[  \partial_m , \Bbb A^m ] $. For any of the gauge-covariant
choices $\Bbb K \rightarrow \{\nabla_\alpha , \nabla_m,\Bbb W^\alpha,\Bbb F^{mn}\}$,
the last term of \boxone\ can be converted to quadratic expressions in the non-linear
fields using \DiracW\ and
\eqnn\cors
$$\eqalignno{
[\nabla_m, \Bbb F^{mp} ]&= \g^p_{\alpha \beta} \{ \Bbb W^\alpha, \Bbb W^\beta \}   \cr
[ \nabla_m, \Bbb W^{m\alpha}]&=  {1\over 2} [\Bbb F_{mn} , (\g^{mn} \Bbb W)^\alpha] &\cors \cr
[\nabla_m , \Bbb F^{m|pq}] &= 2 [\Bbb F^{pn} , \Bbb F_n{}^q]
+ 4 \{ (\Bbb W^{[m} \g^{n]})_\alpha , \Bbb W^{\alpha} \} 
 \ .
}$$
Upon inserting \cors\ into \boxone, one can reproduce the wave
equations \boxAalpha, \boxAm, \altboxW\ and \altboxF\
from $\Bbb K \rightarrow \{\nabla_\alpha , \nabla_m,\Bbb W^\alpha,  \Bbb F^{mn}\}$.

\subsec Generating series of gauge transformations

In general, the non-linear gauge transformations \NLgauge\ are a symmetry of the non-linear
SYM equations of motion \SYMeom\ for any Lie algebra-valued gauge parameter
$\OOO$ with generating series,
\eqn\gaugepar{
\OOO = \sum_{p=1}^{\infty} \sum_{i_1,i_2,\dots,i_p}  \Omega_{i_1 i_2 \ldots i_p}
t^{i_1} t^{i_2} \ldots t^{i_p} \,,
\qquad  \Omega_{A\shuffle B} = 0 \ \forall A,B \neq \emptyset \,.
}
In the remainder of this work we will exploit the effects of 
different gauge parameters $\Omega_P$.
One particular choice to be discussed in the next subsection
efficiently encodes the multiparticle response to linearized gauge variations \lingauge, 
possibly for several external legs.
But more importantly, the multiparticle gauge freedom parameterized
by $\Omega_P$
can be exploited as a tool~to:
\medskip
\item{1.} Find a representation of multiparticle superfields which
manifestly obey generalized Lie symmetries, so-called {\it BCJ
representations} discussed in section~\labBCJrep.
\item{2.} Considerably simplify the theta-expansions of multiparticle
superfields as discussed in section~\labSimpletheta.
\item{3.} Find a multiparticle representation which combines both features above.
\medskip
\noindent
The benefits for scattering amplitudes are sketched in section 5, and the tree-level applications 
are deepened in \otherpaper.

\subsubsec Generating series of polarization shifts

Standard linearized gauge variations of the form $\d_{\cG}A^i_m = k^i_m
\cG_i$ with scalar parameter $\cG_i=e^{k_ix}$ induce multiparticle transformations of the 
Berends--Giele currents by their recursive construction, see \BGdef\ to \cFmn.
They effectively shift gluon polarizations $e^i_m$ by $k^i_m$ and do not affect the transversality $(k_i \cdot A_i)=0$, 
hence, they cannot alter the property $k_P^m {\cal A}_m^P=0$ at any multiparticle level and 
preserve Lorentz gauge \Lorentz. The resulting condition $[ \partial^m,\delta_{\cG } \Bbb A_m]=0$
applied to $\d_\cG \Bbb A_m = [\p_m, \Bbb G] - [\Bbb A_m, \Bbb G]$ (see \NLgauge)
yields a recursion for the multiparticle components of $\Bbb G$,
\eqn\gaugarec{
\Box \Bbb G = 
\big[ \Bbb A_m, [\partial^m ,\Bbb G] \big] \ , \ \ \ \ {\cal G}_P = -{1 \over2s_{P}} \sum_{XY=P}  \big[ {\cal G}_X (k^X \cdot {\cal A}^Y)
-(X\leftrightarrow Y) \big] 
 \ .
}
This formula generalizes the transformations of multiparticle fields discussed
in \cohomology. In that reference the
single-particle initial conditions for the recursion in \gaugarec\ were
specialized to ${\cal G}_i = \delta_{i,1} e^{k_1 x}$; only the gluon
polarization of particle $i=1$ is shifted. Note that \gaugarec\ with several
non-vanishing ${\cal G}_i$ in the initial conditions allows to address simultaneous
shifts of multiple polarization vectors $e^m_i$ by the corresponding $k_i^m$. One can
show that \gaugarec\ is the supersymmetric generalization of the complicated-looking
formula (2.24) of \BerendsZN, highlighting the benefits of the superspace
approach to the Berends--Giele currents adopted here.

\newsec Non-linear superfields and Berends--Giele currents in BCJ gauge\par
\seclab\labBCJrep

\noindent In a previous paper, supersymmetric Berends--Giele currents were
constructed in a totally different fashion \EOMBBs. Starting with a {\it local} representation of multiparticle
superfields
\eqn\localdef{
K_P \in \{ A_\a^P, A_m^P, W_P^\a, F^P_{mn}\}\,,
}
redefinitions were employed in order to enforce the symmetries of nested
commutators $[[t^1,t^2],t^3]$ in a Lie algebra such as $K_{123} + K_{231} + K_{312}= 0$.
Their Berends--Giele currents were constructed by adjoining propagators, i.e.~inverse Mandelstam 
invariants \multmand, to Lie symmetry-satisfying numerators, following an
intuitive mapping to cubic graphs compatible with the ordering of the external
legs. Despite their different construction, the Berends--Giele currents $\cK_P^{\rm BCJ}$ of \EOMBBs\ or those
in the {\it Lorentz gauge} $\cK_P^{\rm L}\equiv \cK_P$ constructed in the previous section give rise to identical tree-level
amplitudes. As verified below up to multiplicity five, these different currents are
in fact related by a non-linear gauge transformation and are therefore equivalent.
As indicated by the superscript in $\cK_P^{\rm BCJ}$, the constituents $K_{12\ldots p}$ of the Berends--Giele currents
in \EOMBBs\ have the symmetries suggested by the BCJ duality between color and kinematics \BCJ. Accordingly, the
currents $\cK_P^{\rm BCJ}$ are said to be in {\it BCJ gauge}.

\subsec Recursive definition of local superfields in Lorentz gauge

The definition of local superfields $\hat K_{[P,Q]}$ in Lorentz gauge\foot{Starting from rank four, the superfields denoted by $\{\hat A^{P}_\a,\hat A^{P}_m, \hat W_{P}^\a , \hat F^{mn}_P\}$ in this work and \EOMBBs\ do not match.} is given by
\eqnn\hatAalpha
\eqnn\hatAm
\eqnn\hatWalpha
$$\eqalignno{
\hat A^{[P,Q]}_\a &= - \half\bigl[  \hat A^{P}_\a (k^{P}\cdot  \hat A^Q)
+  \hat A^{P}_m (\g^m \hat W^Q)_\a - (P\leftrightarrow Q)\bigr]  &\hatAalpha\cr
\hat A^{[P,Q]}_m &= - \half\bigl[ \hat A^{P}_m (k^P\cdot \hat A^{Q}) + \hat A^{P}_n
\hat F^Q_{mn}
- ( \hat W^{P}\g_m \hat W^Q)
- (P \leftrightarrow Q)\bigr] &\hatAm\cr
\hat W_{[P,Q]}^\a &=  \half  (k_P^m + k_Q^m) \gamma_m^{\alpha \beta} \bigl[ \hat A_P^n (\gamma_n \hat W_Q)_\beta 
- (P \leftrightarrow Q) \bigr]\,,&\hatWalpha\cr
}
$$
it amounts to picking up the numerator on top of various inverse $s_{X}$
in the recursions \cAalpha\ to \cWalpha\ for Berends--Giele currents. We
will often use a simplified notation for brackets $[P,Q]$ when one of
$P, Q$ is of single-particle type,
\eqn\nobracket{
\hat K_{12\dots p} \equiv \hat K_{[12\ldots p-1,p]} \ .
}
In this topology, the field-strength\foot{Field-strenghts $\hat F^{mn}_{[P,Q]}$ of more general
topologies beyond \nobracket\ such as $\hat F^{mn}_{[12,34]}$ can be
addressed along the lines of \altW.
}
appearing above is given by
\eqn\hatFmn{
\hat F^{12 \ldots p}_{mn} \equiv k^{12 \ldots p}_m \hat A^{12 \ldots p}_n - k^{12 \ldots p}_n \hat A^{12 \ldots p}_m
+ \sum_{j=2}^{p} \sum_{\d \in P(\beta_j)}   \! \! \! \! (k_{12 \ldots j-1}\cdot k_j)\,
\hat A_{[n}^{12{\ldots} j-1,\{\d\}}\,  \hat A_{m]}^{j, \{\beta_j \backslash \d \} }
\,,}
where $\b_j \equiv \{j+1, j+2,{\ldots},p\}$ and $P(\beta_j)$ denotes its power set.

\subsec Review of generalized Lie symmetries for multiparticle superfields

The approach of \EOMBBs\ to Berends--Giele currents in BCJ gauge ${\cal
K}_P^{\rm{BCJ}}$ is based on local superfields $K_{12\ldots p}$
satisfying all generalized Lie symmetries $\lie_k$ up to $k=p$,
\eqnn\BRSTsym
$$\displaylines{
\openup3pt
\lie_k\circ K_{12 \ldots p} =0, \quad k=2, \ldots,p
\hfil\BRSTsym\hfilneg\cr
\eqalign{
\lie_{k=2n+1}\colon\quad&{} K_{12\ldots n+1[n+2[\ldots [2n-1[2n,2n+1]] \ldots ]]}
- K_{2n+1\ldots n+2[n+1[\ldots [3[21]] \ldots ]]}  = 0 \cr
\lie_{k=2n}\colon\quad&{} K_{12\ldots n[n+1[\ldots [2n-2[2n-1,2n]] \ldots ]]}
+ K_{2n\ldots n+1[n[\ldots [3[21]] \ldots ]]}  = 0 \ .\cr
}}$$
For example, 
\eqnn\Lietwo
$$\displaylines{\lie_2\circ K_{12} = K_{12} + K_{21} = 0\,,\ \ \ \ \ \
\lie_3\circ K_{123} = K_{123} + K_{231} + K_{321} = 0 \hfil\Lietwo\hfilneg\cr
\lie_4\circ K_{1234} = K_{1234} - K_{1243} + K_{3412} - K_{3421}=0\,,
}$$
and so forth. These symmetries
leave $(p-1)!$ independent permutations of
$K_{12\ldots p}$ and are also obeyed by nested commutators
$[\ldots [[ t^1, t^2 ], t^3 ],\ldots ,t^p]$ 
and the color factors in
\eqn\Liethree{
K_{12\ldots p}\ \longleftrightarrow f^{1 2 a_3} f^{a_3 3 a_4} f^{a_4 4 a_5}
\ldots f^{a_p p a_{p+1}} \,.
}
Therefore the local superfields $K_P$ admit the following diagrammatic
interpretation:
\medskip
\centerline{{\epsfxsize=.46\hsize\epsfbox{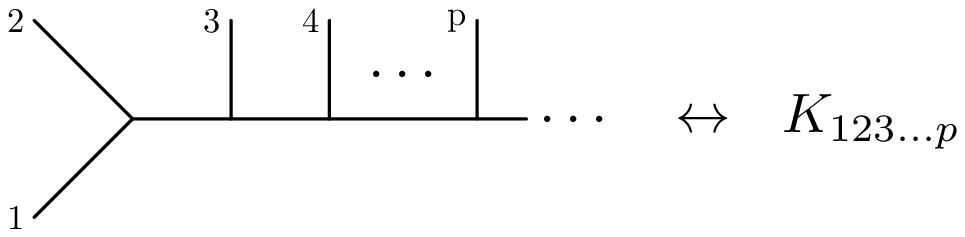}}}

\subsec Recursive definition of local superfields in BCJ gauge

The recursively defined superfields $\hat K_{12\ldots p}$ in \hatAalpha\ to
\hatFmn\ do not yet satisfy the Lie symmetries \BRSTsym. However,
this can be compensated by redefinitions
$K_{12\ldots p}= \hat K_{12\ldots p}+\ldots$ via superfields
$\hat H_{12 \ldots p}\equiv \hat H_{[12 \ldots p-1,p]}$ which
amount to a non-linear gauge transformation of their corresponding
generating series. Starting from $\hat H_{i} = \hat H_{ij} = 0$,
the superfields $\hat H_{12 \ldots p}$ at multiplicity $p$
enter through the following recursive system of equations \EOMBBs\
\eqnn\Hpdef
$$\eqalignno{
K_{[12 \ldots p-1,p]} &\equiv \hat K_{[12 \ldots p-1,p]} -
\sum_{j=2}^{p}\sum_{\d \in P(\b_j)}\!\! (k^{1\ldots {j-1}}\cdot k^j)
\bigl[ \hat H_{1\ldots j-1,\{\d\}}\;  \hat K_{j,\{\b_j\backslash \d\}}
- (1\ldots j-1\leftrightarrow j)\bigr] \cr
& \ \ \ \ \ \ \ - \cases{ D_\alpha \hat H_{[12\ldots p-1,p]} &: \ $K= A_\alpha$\cr
k_{12\ldots p}^m \hat H_{[12\ldots p-1,p]} &: \ $K= A^m$ \cr
0 &: \ $K= W^\alpha$} 
&\Hpdef
}$$
and will be introduced separately in the next subsection.

The redefinitions in \Hpdef\ have been originally designed in a two-step procedure
which yields the expressions for $\hat H_{12\ldots p}$
in a constructive manner\foot{
As discussed in \EOMBBs, an intermediate step of the redefinition procedure
gives rise to redefined superfields
$A'^{m}_{12\ldots p}$ which determine the definition of $H_{[12\ldots p-1, p]}$ via
$\lie_p\circ A'^{m}_{[12 \ldots p-1,p]} \equiv p\, k^m_{12 \ldots p} H_{[12\ldots p-1,p]}$.
For this definition to work,
the overall momentum $k^m_{12 \ldots p}$ must factorize in the
sum dictated by $\lie_p\circ A'^{m}_{[12 \ldots p-1,p]}$, providing a
strong consistency check of the setup.
The relation between $H_{12\ldots p}$ and $\hat H_{12\ldots p}$ will be
given in \Hhatrecurs.
} \EOMBBs. As a result, the superfields $K_{12\ldots p}$ defined by \Hpdef\ as well as
\eqn\FinalDefs{
F^{12 \ldots p}_{mn} \equiv k^{12 \ldots p}_m A^{12 \ldots p}_n - k^{12 \ldots p}_n A^{12 \ldots p}_m
+ \sum_{j=2}^{p} \sum_{\d \in P(\beta_j)}   \! \! \! \! (k_{12 \ldots j-1}\cdot k_j)\,
A_{[n}^{12{\ldots} j-1,\{\d\}}\,  A_{m]}^{j, \{\beta_j \backslash \d \} }
}
satisfy all the Lie symmetries $\lie_2,\lie_3,\ldots$ in \BRSTsym\ up to and including $\lie_p$.
For example, since $\hat H_i = \hat H_{ij}=0$, the definitions in \Hpdef\ yield
\eqn\rankONETWO{
K_1 = \hat K_{1} \ , \ \ \    \ \ \ K_{12} = \hat K_{12} \ , \ \ \   \ \ \ \forall \ K\in \{ A_\alpha, A^m ,W^\alpha, F^{mn}\}  \ ,
}
and the first non-trivial redefinition occurs at multiplicity three with
\eqn\rankThreeEx{
 A^{123}_\alpha = \hat A^{[12,3]}_\alpha - D_\alpha \hat H_{[12,3]} \ , \ \ \ \ 
 A^m_{123} = \hat A^m_{[12,3]} -  k^m_{123} \hat H_{[12,3]} \ , \ \ \ \ 
 W^\alpha_{123} = \hat W^\alpha_{[12,3]}  \ .
}
A rank-four sample of the redefinitions \Hpdef\ is provided by
\eqnn\rkfoursample
$$\eqalignno{
A^m_{1234} &= \hat A^{m}_{[123,4]} -(k^{123}\cdot k^4) \hat H_{[12,3]}A_4^m
-(k^{12}\cdot k^3) \hat H_{[12,4]}A_3^m \cr
&
-(k^1\cdot k^2)\big( \hat H_{[13,4]}A_2^m - \hat H_{[23,4]}A_1^m \big)
- k^m_{1234}\hat H_{[123,4]}  \ . & \rkfoursample
}$$

\subsec Explicit form of the redefinitions $\hat H$

\noindent
One can show that expressions for $\hat H_{[12\ldots p-1,p]}$ can be conveniently
summarized by
\eqnn\Hhatrecurs
\eqnn\HtwoOneOne
$$\eqalignno{
\hat H_{[A,B]}&\equiv H_{[A,B]} 
- \half \big[\hat H_{A}(k_{A}\cdot A_B) - (A\leftrightarrow B)\big] &\Hhatrecurs
\cr
H'_{A,B,C} &\equiv H_{A,B,C}
+ {1\over 2}\big[H_{[A,B]}(k_{AB}\cdot A_C) + {\rm cyclic}(A,B,C)\big]\,, &\HtwoOneOne
}$$
with the central building block
\eqn\HABCdef{
H_{A,B,C} \equiv -{1\over 4}A^m_A A^n_B F^{mn}_C
+ \half (W_A\g_m W_B)A_C^m + {\rm cyclic}(A,B,C)\,.
}
In particular, the redefinitions up to multiplicity five
are captured by
\eqnn\topologies
$$\eqalignno{
H_{[12,3]}  &={1\over 3} H_{1,2,3}\cr
H_{[123,4]} &={1\over 4}\big(H'_{12,3,4} + H'_{34,1,2}\big) \cr
H_{[12,34]} &= {1\over 4}\big(-2 H'_{12,3,4} + 2H'_{34,1,2}\big) &\topologies\cr
H_{[1234,5]} &= {1\over 5}\big(H'_{123,4,5} - H'_{543,2,1} + H'_{12,3,45}\big)\cr
H_{[123,45]} &= {1\over 5}\big(-3H'_{123,4,5} - 2H'_{543,2,1} + 2H'_{12,3,45}\big) \ .
}$$
The treatment and significance of the additional topologies $H_{[12,34]}$ and
$H_{[123,45]}$ is explained around \extratop\ and in appendix \BCJapp. Higher-rank
versions of $H_P$ are under investigation, and it would be
interesting to extend the simple expressions in \topologies\ to arbitrary
multiplicity\foot{Noting that $H_{[12 \ldots p-1,p]}$ here corresponds to
$H_{12 \ldots p}$ from \EOMBBs, the expression of $H_{[123,4]}$
presented in \topologies\ considerably simplifies the
expression of $H_{1234}$ given in the appendix C of \EOMBBs.}.
The expressions above are sufficient to identify the redefinitions up
to and including multiplicity five as originating from a non-linear gauge
transformation.

It is worth mentioning a remarkable feature of $H_{A,B,C}$ in \HABCdef:
Upgrading the polarization vectors and spinors in
the color-ordered SYM three-point amplitude at tree level,
\eqn\AYMthree{
A^{\rm SYM}(1,2,3) = -{1\over 2}  e^m_1 e^n_2 f^{mn}_3
+ (\chi_1\g_m \chi_2)e_3^m + {\rm cyclic}(1,2,3)\,,
}
to superfields according to $e^m_i \rightarrow A^m_i(\t)$,
$\chi^\a_i \rightarrow W^\a_i(\t)$ and $f^{mn}_i=k_i^{[m} e_i^{n]} \rightarrow F_i^{mn}(\t)$, the amplitude \AYMthree\ can be
rewritten as
\eqn\AYMsup{
A^{\rm SYM}(1,2,3) = 2 H_{1,2,3}(\t=0) \,.
}

\subsec Supersymmetric Berends--Giele currents in BCJ gauge

In this section, we will justify the terminology of Lorentz and BCJ gauge for
the representations 
${\cal K}^{\rm L}_{P}$ and ${\cal K}^{\rm BCJ}_{P}$ of Berends--Giele currents. 
It will be verified up to multiplicity five that they are indeed related by
a non-linear gauge transformation, i.e.
\eqn\toBCJ{
\Bbb A_m^{{\rm BCJ}} =\Bbb A_m^{{\rm L}} - [\partial_m,\Bbb H]
+ [\Bbb A_m^{{\rm L}}, \Bbb H] \ ,
}
translating into
\eqn\toshow{
\cA^{m,\rm BCJ}_{P} =\cA^{m,\rm L}_{P} - k^m_{P}\cH_{P} + \sum_{XY=P} ( \cA^{m,\rm L}_{X} {\cal H}_Y - \cA^{m,\rm L}_{Y} {\cal H}_X) \ .
}
Clearly, \toBCJ\ is a special case of a non-linear gauge transformation \NLgauge\ with
$\OOO \rightarrow - \Bbb H$. The generating series of gauge parameters
\eqn\Hseries{
\Bbb H \equiv \sum_{i_1,i_2,i_3} {\cal H}_{i_1 i_2 i_3} t^{i_1} t^{i_2} t^{i_3} +
\sum_{i_1,i_2,i_3,i_4} {\cal H}_{i_1 i_2 i_3 i_4} t^{i_1} t^{i_2} t^{i_3} t^{i_4}+\cdots 
}
is built from Berends--Giele currents $\cH_P$ of the superfields $\hat H_{[A,B]}$.
As before, the Berends--Giele symmetry
${\cal H}_{A \shuffle B}=0$ implies Lie algebra-valuedness of the series \Hseries\
\ree. Of course, the same $\Bbb H$ describes the transformation of the remaining
series $\Bbb A_\alpha$, $\Bbb W^\a$, $\Bbb F^{mn}$, see \NLgauge. We will focus on
the transformation between the currents $\cA^{m,\rm BCJ}_{P}$ and $\cA^{m,\rm
L}_{P}$ of the vector potential since the remaining superfields follow the same or simpler lines.

In the following discussion we will construct Berends--Giele currents
up to rank four using the mapping between planar binary trees and nested brackets \EOMBBs, 
see appendix \BCJapp\ for rank five.
By \rankONETWO, the two gauge choices are identical at multiplicities one and two,
\eqn\rkonetwo{
{\cal K}_{1}^{{\rm BCJ}}= {\cal K}_{1}^{{\rm L}} \ , \ \ \ \ \ \ 
{\cal K}_{12}^{{\rm BCJ}}= {\cal K}_{12}^{{\rm L}} \, ,
}
reflecting the vanishing of the simplest redefinitions,
\eqn\Honetwo{
\hat H_1 = \hat H_{12} = 0 \ \ \ \Rightarrow \ \ \ {\cal H}_1 = {\cal H}_{12} = 0 \ ,
}
and justifying the absence of single-particle and two-particle contributions in the series \Hseries.

\subsubsec Rank three

At multiplicity three, the two binary trees displayed in \figPlanarTrees\
lead to
\eqn\BGthree{
\cK^{\rm BCJ}_{123} = 
{K_{[12,3]}\over s_{12}s_{123}}
+ {K_{[1,23]}\over s_{23}s_{123}},\qquad
\cK^{\rm L}_{123} = 
{\hat K_{[12,3]}\over s_{12}s_{123}}
+ {\hat K_{[1,23]}\over s_{23}s_{123}} \,,
}
with $\hat K_{[P,Q]} =-\hat K_{[Q,P]}$ from \hatAalpha\ to \hatWalpha.
Hence, the relation \rankThreeEx\ between the local superfields in the two gauges
is sufficient to determine the corresponding relation between their
Berends--Giele currents. For example,
$A^{m}_{[12,3]} = \hat A^{m}_{[12,3]} - k^m_{123}\hat H_{[12,3]}$ implies
that
\eqn\BCJLorentzThree{
\cA^{m,\rm BCJ}_{123} =\cA^{m,\rm L}_{123} - k^m_{123}\cH_{123}, \qquad
\cH_{123} =
{\hat H_{[12,3]}\over s_{12}s_{123}}
+ {\hat H_{[1,23]}\over s_{23}s_{123}} \ ,
}
where \Honetwo\ allows to restore a vanishing deconcatenation
term $0= \cA^{m,\rm L}_{1} {\cal H}_{23}  +\cA^{m,\rm L}_{12} {\cal H}_{3}
-\cA^{m,\rm L}_{23} {\cal H}_{1}  -\cA^{m,\rm L}_{3} {\cal H}_{12}$ and to verify \toshow\ at $P=123$.

\ifig\figPlanarTrees{The planar binary trees used to define $\cK_{123}$ and $\cK_{1234}$.}
{\vbox{\centerline{\epsfxsize=.40\hsize\epsfbox{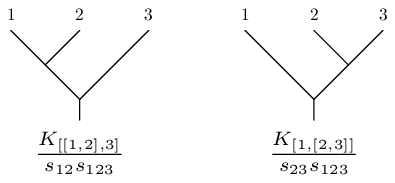}}\vskip6pt
\centerline{\epsfxsize=.92\hsize\epsfbox{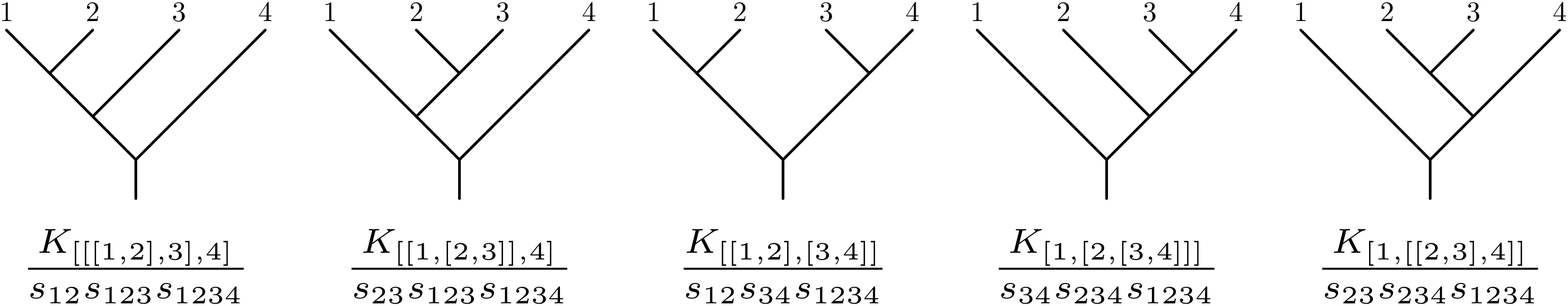}}
}}

\subsubsec Rank four

\noindent
Similar calculations at multiplicity four lead to the relation
\eqn\BCJLorentzFour{
\cA^{m,\rm BCJ}_{1234} =\cA^{m,\rm L}_{1234} - k^m_{1234}\cH_{1234}
+ \cA^m_1 \cH_{234} - \cA^m_4 \cH_{123}
}
required by \toshow, where \Honetwo\ identifies the last two terms on the right-hand side as a perfect 
deconcatenation $\sum_{XY=1234} (\cA^{m,\rm L}_{X} {\cal H}_{Y}-\cA^{m,\rm L}_{Y} {\cal H}_{X})$. 
The Berends--Giele currents comprise the five binary trees depicted in \figPlanarTrees,
\eqnn\BGfours
$$\eqalignno{
\cA^{m,\rm BCJ}_{1234} &= {1 \over s_{1234}}
\Big( {A^{m }_{[123,4]}\over s_{12}s_{123} }
 + {A^{m }_{[321,4]}\over s_{23}s_{123} }
 + {A^{m }_{[12,34]}\over s_{12}s_{34} }
 + {A^{m }_{[342,1]}\over s_{34}s_{234} }
 + {A^{m }_{[324,1]}\over s_{23}s_{234} }
 \Big) \cr
\cA^{m,\rm L}_{1234} &= {1 \over s_{1234}}
\Big( {\hat A^{m }_{[123,4]}\over s_{12}s_{123} }
 + {\hat A^{m }_{[321,4]}\over s_{23}s_{123} }
 + {\hat A^{m }_{[12,34]}\over s_{12}s_{34} }
 + {\hat A^{m }_{[342,1]}\over s_{34}s_{234} }
 + {\hat A^{m }_{[324,1]}\over s_{23}s_{234} }
 \Big)  &\BGfours \cr
\cH_{1234} &= {1 \over s_{1234}} \Big( {\hat H_{[123,4]}\over s_{12}s_{123} }
 + {\hat H_{[321,4]}\over s_{23}s_{123} }
 + {\hat H_{[12,34]}\over s_{12}s_{34} }
 + {\hat H_{[342,1]}\over s_{34}s_{234} }
 + {\hat H_{[324,1]}\over s_{23}s_{234} }
 \Big)\,,
}$$
where four of the five numerators in \BGfours\ belong to the topology of \rkfoursample. 
However, the third term representing the middle diagram in \figPlanarTrees\ follows the separate conversion rule
\eqnn\extratop
$$\eqalignno{
&A^{m}_{[12,34]}  = \hat A^{m}_{[12,34]} - k^m_{1234} \hat H_{[12,34]} &\extratop \cr
& \ \ \ \ +(k^1\cdot k^2)\big( \hat H_{[13,4]}A_2^m - \hat H_{[23,4]}A_1^m \big)+(k^3\cdot k^4)\big( \hat H_{[12,4]}A_3^m - \hat H_{[12,3]}A_4^m \big)\cr
}$$
between Lorentz gauge and BCJ gauge. As a defining property of BCJ gauge, the left-hand
side can be expressed in terms of the basic topology \Hpdef\ via $A^{m}_{[12,34]} = A^m_{1234}-A^m_{1243}$. 
The new topology $\hat H_{[12,34]}$ of redefining fields (see \cohomology) 
is determined by \extratop\ whose solution can be found in \topologies.

Upon insertion into \BGfours, contributions of the form $\hat H_{[12,3]} A_4^m$ in
\rkfoursample\ and \extratop\ conspire to the desired deconcatenation term in
\BCJLorentzFour, verifying the mediation of a non-linear gauge transformation
between $\cA^{m,\rm BCJ}_{1234}$ and $\cA^{m,\rm L}_{1234}$. The analogous analysis
of the gauge transformation at multiplicity five is relegated to appendix~\BCJapp.

\newsec Theta-expansions in Harnad--Shnider gauge\par
\seclab\labSimpletheta

\noindent In the last section we have identified a particular gauge transformation $\Bbb
H$ which relates the Berends--Giele currents in the BCJ gauge to their counterparts
in the Lorentz gauge. Similarly, we will now construct another gauge transformation
\eqn\Lseries{
\Bbb L \equiv
\sum_{i_1,i_2} {\cal L}_{i_1 i_2 } t^{i_1} t^{i_2}  +
 \sum_{i_1,i_2,i_3} {\cal L}_{i_1 i_2 i_3} t^{i_1} t^{i_2} t^{i_3}  + \cdots
}
whose expansion starts at multiplicity two and
is designed to simplify the theta-expansions of the multiparticle
superfields.

\subsec Generating series of Harnad--Shnider gauge variations

A convenient gauge choice to expand the superfields of ten-dimensional
SYM in theta is the Harnad--Shnider (HS) gauge \HarnadBC,
\eqn\HSgauge{
\theta^\alpha \Bbb A^{{\rm HS}}_\alpha = 0 \,.
}
At the linearized level, the gauge $\t^\a A_\a^i =0$ has been used in
\PolicastroVT\ to obtain the theta-expansions of the single-particle superfields
to arbitrary order. However, the recursive definition \cAalpha\ of multiparticle
Berends--Giele currents ${\cal A}_\alpha^P$ in
Lorentz gauge does not preserve linearized HS gauge, e.g.
\eqn\noHS{
\t^\a A_\a^i =0  \ \ \Rightarrow \ \
\theta^\alpha {\cal A}_\alpha^{12}={1\over 2s_{12}}\big[ A_m^2 (\theta \gamma^m W_1) - (1\leftrightarrow 2) \big] \neq 0 
\ .
}
Still, there is a non-linear gauge
transformation $\Bbb L$ which brings the currents from Lorentz gauge into HS gauge via
\eqn\toHS{
\Bbb A^{{\rm HS}}_\alpha = \Bbb A^{{\rm L}}_\alpha - [D_\alpha ,\Bbb L] + [\Bbb
A^{{\rm L}}_\alpha , \Bbb L]\,.
}
It can determined recursively by contracting with $\theta^\alpha$:
\eqn\getL{
[{\cal D},\Bbb L] = \theta^\alpha \Bbb A^{{\rm L}}_\alpha + [\theta^\alpha \Bbb
A^{{\rm L}}_\alpha , \Bbb L]  \,,
}
where the Euler operator
\eqn\Euler{
{\cal D} \equiv \theta^\alpha D_\alpha = \theta^\alpha {\partial \over \partial \theta^\alpha}
}
weights the $k^{\rm th}$ order in $\theta$ by a factor of $k$.
At the level of multiparticle components in \Lseries, this translates into
\eqn\recL{
{\cal D}{\cal L}_P = \theta^\alpha {\cal A}^{P}_\alpha + \sum_{XY=P} \big(  \theta^\alpha {\cal A}^{X}_\alpha  {\cal L}_Y - 
 \theta^\alpha {\cal A}^{Y}_\alpha  {\cal L}_X \big) \ ,
}
where the Berends--Giele currents ${\cal L}_X, {\cal L}_Y$ on the right hand side have lower multiplicity
than ${\cal L}_P$ on the left hand side. Hence, \recL\ is a recursion w.r.t.
multiplicity in the Lie-series expansion \Lseries. The currents ${\cal A}^{P}_\alpha$ are
understood to follow the Lorentz-gauge setup in \BGdef\ to \cFmn. Using
$\theta^\alpha A_\alpha^i={\cal L}_i=0$ at the linearized level, we have for instance
\eqn\recLtwo{
{\cal D}{\cal L}_{12} = \t^\a {\cal A}^{12}_\a\,,\quad
{\cal D}{\cal L}_{123} = \t^\a {\cal A}^{123}_\a\,,\quad
{\cal D}{\cal L}_{1234} = \t^\alpha {\cal A}^{1234}_\a
+ \t^\a {\cal A}^{12}_\alpha{\cal L}_{34} -\t^\alpha {\cal A}^{34}_\alpha{\cal L}_{12}\,.
}
By imposing $\Bbb L(\t=0) = 0$, we arrive at explicit theta-expansions such as
\eqnn\Ltwo
$$\eqalignno{
{\cal L}_{12} &= {1\over 2s_{12}}\Big((\t \gamma_m \chi_1)  e_2^m + {1\over 8} (\t \gamma_{mnp} \t) e_1^m f_2^{np}   \cr
& \ \ \ \ \ \ \ + {1\over 12}(\t \gamma_{mnp}\t) (\t \gamma^m \chi_1 ) k_{12}^n
e_2^p - (1\leftrightarrow 2)+ \cdots \Big) e^{k_{12} x}\,, &\Ltwo
}$$
with terms of order $\t^{\geq 4}$ in the ellipsis and analogous expressions for ${\cal L}_{12\ldots p}$ at $p \geq 3$.

\subsec Multiparticle theta-expansions in Harnad--Shnider gauge
\par\subseclab\towHS

\noindent
The theta-expansion of non-linear fields in HS gauge \HSgauge\ can be
elegantly captured by means of higher mass dimension superfields \MafraGIA,
\eqn\highmass{
\Bbb W^{m_1\dots m_k\alpha}\equiv[\nabla^{m_1},\Bbb W^{m_2\dots m_k\alpha}] \ , \ \ \ \
\Bbb F^{m_1\dots m_k\vert pq}\equiv[\nabla^{m_1},\Bbb F^{m_2\dots m_k\vert pq}] \ ,
}
subject to non-linear gauge transformations \MafraGIA
\eqn\highgauge{
 \delta_{\OOO }   \Bbb W^{m_1\ldots m_k \alpha} =  \big[\OOO, \Bbb W^{m_1\ldots m_k \alpha}  \big] 
 ,\quad \delta_{\OOO }  \Bbb F^{m_1\ldots m_k|pq} =  \big[\OOO, \Bbb F^{m_1\ldots m_k|pq}  \big] \ .
}
In the subsequent, we assume that the superfields have been brought to HS
gauge via \NLgauge\ through the transformation $\OOO \rightarrow \Bbb L$
constructed from \recL. For ease of notation, the accompanying $^{ {\rm HS}}$
superscripts as in \toHS\ will henceforth be suppressed. Contracting the non-linear equations
of motion \SYMeom\ with $\theta^\alpha$ yields \HarnadBC
\eqnn\SYMeomt
$$\eqalignno{
\big( {\cal D}+1\big)\Bbb A_{\beta}&=(\theta\gamma^m)_\beta\Bbb A_m \ , \ \ \ \ \ \ \ \ \ \ 
{\cal D}\Bbb A_m=(\theta\gamma_m\Bbb W) &\SYMeomt \cr
{\cal D}\Bbb W^\beta&={1\over4}(\theta\gamma^{mn})^\beta\Bbb F_{mn} \ , \ \ \ \
{\cal D}\Bbb F^{mn}=-(\Bbb W^{[m}\gamma^{n]}\theta)
}
$$
by virtue of HS gauge. This can be used to reconstruct the entire theta-expansion
of any superfield from their zeroth orders $\Bbb K(\theta=0)$ \HarnadBC,
\eqnn\thetaex
$$\eqalignno{
[\Bbb A_{\alpha}]_k&={1\over k+1}(\theta\gamma^m)_\alpha[\Bbb A_m]_{k-1} \ , \ \ \ \
[\Bbb A_m]_k={1\over k}(\theta\gamma_m[\Bbb W]_{k-1}) &\thetaex  \cr
[\Bbb W^\alpha]_k&={1\over4k}(\theta\gamma^{mn})^\alpha[\Bbb F_{mn}]_{k-1} \ , \ \ \ \
[\Bbb F^{mn}]_k=-{1\over k}([\Bbb W^{[m}]_{k-1}\gamma^{n]}\theta) \ ,   
}
$$
where the notation $[\ldots]_k$ instructs to only keep terms of order $(\theta)^k$ of
the enclosed superfields.
The analogous expressions for superfields at higher mass dimensions are
\eqnn\moretheta
$$\eqalignno{
[\Bbb W_m^{\alpha}]_k&={1\over k}\bigg\{{1\over4}(\theta\gamma^{pq})^\alpha[\Bbb F_{m\vert pq}]_{k-1} -  (\theta \gamma_m)_\beta
\sum_{l=0}^{k-1}\{ [\Bbb W^\beta]_l,[\Bbb W^{\alpha}]_{k-l-1}\}\bigg\}\cr
[\Bbb F^{m\vert pq}]_k&=-{1\over k}\bigg\{([\Bbb W^{m[p}]_{k-1}\gamma^{q]}\theta) + (\theta \gamma^m)_\alpha \sum_{l=0}^{k-1}[ [\Bbb W^\alpha]_l,[\Bbb F^{ pq}]_{k-l-1}]\bigg\}&\moretheta \cr
[\Bbb W_{mn}^{\alpha}]_k&={1\over k}\bigg\{{1\over4}(\theta\gamma^{pq})^\alpha[\Bbb F_{mn\vert pq}]_{k-1}+(\theta\gamma_m)_\beta
\sum_{l=0}^{k-1}     \{ [\Bbb W^\beta]_l,[\Bbb W_n^{\alpha}]_{k-l-1}\}      \cr
& + (\theta\gamma_n)_\beta \sum_{l=0}^{k-1} \Big(  \{ [\Bbb W^{\beta}_{m}]_l,[\Bbb W^{\alpha}]_{k-l-1}\}
+ \{ [\Bbb W^\beta]_l,[\Bbb W_m^{\alpha}]_{k-l-1}\}
\Big)
\bigg\} \ ,
}
$$
see \MafraGIA\ for the underlying equations of motion and \mosttheta\ for generalizations to higher mass dimension. 

\subsubsec The component wavefunctions

The theta-independent terms $[\Bbb K]_0$ initiate the above recursions in the order of theta, and their multiparticle components $[{\cal K}_P]_0$ at lowest mass dimensions
\eqn\sameBG{
[\cA_P^m]_0 \equiv \ce^m_P e^{k_Px}  \ , \ \ \ \ \ \ 
[\cW^\a_P]_0 \equiv {\cal X}^\alpha_P e^{k_Px}  
}
are shown in \otherpaper\ to supersymmetrize the Berends--Giele currents in \BerendsME, e.g.
\eqnn\twocurr
$$\eqalignno{
s_{12} \ce_{12}^m &= e_2^m (k_2\cdot e_1) - e_1^m (k_1 \cdot e_2)+ {1\over 2}(k_1^m- k_2^m)(e_1 \cdot e_2)
+ (\chi_1 \gamma^m \chi_2) &\twocurr
\cr
s_{12} {\cal X}_{12}^\alpha &= \half k_{12}^m \gamma_m^{\alpha\beta} \big[ e_1^n (\gamma_n \chi_2)_\beta - e_2^n (\gamma_n \chi_1)_\beta \big] 
  \ .
}$$
Note that Lorentz gauge for the superfields ${\cal A}_P^m$ propagates to the component currents $\ce_{P}^m$,
\eqn\transverseBG{
(k_P\cdot \ce_P) = (k_P\cdot [{\cal A}_P]_0)= 0\,,
}
since the transformation towards HS gauge in \getL\ is chosen with $\Bbb L(\theta=0)=0$.

At higher mass dimensions, the wavefunctions in
\eqn\moreBG{
[\cW^{m_1\ldots m_k\alpha}_P]_0 \equiv {\cal X}^{m_1\ldots m_k\alpha}_P e^{k_Px}   \ , \ \ \ \ \ \
[\cF^{m_1\ldots m_k|pq}_P]_0 \equiv \cf^{m_1\ldots m_k|pq}_P e^{k_Px}  
}
with $k=0,1,2,\ldots$ inherit the recursive expressions from \highmass\ such that
\eqnn\compBGzero
$$\eqalignno{
 \cf^{mn}_P  &\equiv   k_P^{m} \ce_P^{n} -  k_P^{n} \ce_P^{m}
- \sum_{XY=P} (\ce_X^{m} \ce_Y^{n} -\ce_X^{n} \ce_Y^{m} ) &\compBGzero \cr
 {\cal X}^{m_1\ldots m_k\alpha}_P&\equiv  k_P^{m_1} {\cal X}^{m_2\ldots m_k|pq}_P
- \sum_{XY=P} (\ce_X^{m_1} {\cal X}^{m_2\ldots m_k\alpha}_Y
- {\cal X}^{m_2\ldots m_k\alpha}_X \ce_Y^{m_1} ) \ , \ \ \ \ k=1,2,\ldots   \cr
\cf^{m_1\ldots m_k|pq}_P &\equiv k_P^{m_1} \cf^{m_2\ldots m_k|pq}_P
- \sum_{XY=P} (\ce_X^{m_1} \cf^{m_2\ldots m_k|pq}_Y
- \cf^{m_2\ldots m_k|pq}_X \ce_Y^{m_1} ) \ , \ \ \ \ \ \ k=1,2,\ldots
\ .
}
$$

\subsubsec The theta-expansion

Using the notation ${\cal K}_P(x,\t)\equiv {\cal K}_P(\t)e^{k_P\cdot x}$ one can show
that the recursions \thetaex\ and \moretheta\ lead
to the following multiparticle theta-expansions,
\eqnn\THEXone
$$\eqalignno{
{\cal A}^P_{\alpha}(\theta)&=
{1\over 2}(\theta\gamma_m)_\alpha \ce^m_P
+{1\over 3}(\theta\gamma_m)_\alpha (\theta\gamma^m{\cal X}_P)
- {1\over32}(\theta\gamma_m)^\alpha(\theta\gamma^{mnp}\theta)\cf^P_{np} &\THEXone\cr
&{}
+ {1\over60}(\t\g_m)_\alpha(\t\g^{mnp}\t)({\cal X}^P_n\g_p\t)
+ {1\over1152}(\t\g_m)_\a(\t\g^{mnp}\t)(\t\g^{pqr}\t)\cf_{P}^{n\vert qr}&\cr
&+ \sum_{XY=P}[\cA^{X,Y}_\a]_5 + \ldots \cr
{\cal A}_P^m(\t)&=
\ce_P^m
+(\t\g^m {\cal X}_P)
-{1\over8}(\t \g^{mpq}\t) \cf_P^{pq}
+{1\over12}(\t\g^{mnp}\t)(\cX_{P}^{n}\g^{p}\t)\cr
&{}+{1\over 192}(\t\g^{m}_{\phantom{m}nr}\t)(\t\g^{r}_{\phantom{r}pq}\t)\cf_P^{n\vert pq}
-{1\over480} (\t\g^{m}_{\phantom{m}nr}\t)(\t\g_{\phantom{r}pq}^{r}\t)(\cX^{np}_{P}\g^{q}\t)\cr
&+ \sum_{XY=P}\Big([\cA_{X,Y}^m]_4 + [\cA_{X,Y}^m]_5\Big)  + \ldots  \cr
\cW_P^\a(\t)&=
\cX^\alpha_P
+{1\over 4}(\t \g^{mn})^{\alpha} \cf^P_{mn}
-{1\over 4}(\t \g_{mn})^{\alpha}(\cX_P^{m}\g^{n}\t)
-{1\over48} (\t\g^{\phantom{m}q}_{m})^\alpha(\t\g_{qnp}\t)\cf^{m\vert np}_P& \cr
&{}+{1\over96}  (\t\g^{\phantom{m}q}_{m})^{\alpha}(\t\g_{qnp}\t)({\cal X}^{mn}_{P}\g^{p}\t)
- {1\over 1920}(\t\g^{\phantom{m}r}_{m})^{\alpha}(\t\g^{\phantom{nr}s}_{nr}\t)(\t\g_{spq}\t)\cf^{mn\vert pq}_{P} \cr
&+ \sum_{XY=P}\Big([\cW_{X,Y}^\a]_3 + [\cW_{X,Y}^\a]_4 + [\cW_{X,Y}^\a]_5\Big)  + \ldots  \cr
{\cal F}_P^{mn}(\t)&=
\cf_P^{mn}
-(\cX_P^{[m}\g^{n]}\t)+{1\over 8}(\t\g_{pq}^{\phantom{pq}[m}\t) \cf_P^{n]\vert pq}
 -{1\over12}(\t\g^{\phantom{pq}[m}_{pq}\t)(\cX_{P}^{n]p}\g^{q}\t)\cr 
&{} -{1\over192}(\t\g^{\phantom{ps}[m}_{ps}\t)\cf^{n]p\vert qr}_P(\t\g^{s}_{\phantom{s}qr}\t)
+ {1\over 480}(\t\g^{[m}_{\phantom{[m}ps}\t)(\cX^{n]pq}_{P}\g^{r}\t)(\t\g^{s}_{\phantom{s}qr}\t)\cr
&+ \sum_{XY=P}\Big([\cF_{X,Y}^{mn}]_2  + [\cF_{X,Y}^{mn}]_3 + [\cF_{X,Y}^{mn}]_4 +
[\cF_{X,Y}^{mn}]_5\Big) + \sum_{XYZ=P} [\cF_{X,Y,Z}^{mn}]_5  + \ldots  \cr
}
$$
with terms of order $\theta^{\geq 6}$ in the ellipsis. The non-linearities of the form
$\sum_{XY=P} [{\cal K}_{X,Y}]_l$ can be traced back to the quadratic
expressions in \moretheta, e.g.
\eqnn\appFa
$$\eqalignno{
[{\cal A}_\alpha^{X,Y}]_5&=\pplus
{1\over144}(\t\g_m)_\alpha(\t\g^{mnp}\t)({\cal X}^X\g_n\t)({\cal X}^Y\g_p\t) &\appFa \cr
%
[{\cal A}_{X,Y}^m]_4 &=\pplus {1\over24} (\t\g^{m}_{\phantom{m}np}\t) ({\cal X}^{X}\g^{n}\t) ({\cal X}^{Y}\g^{p}\t)   \cr
[{\cal W}_{X,Y}^\alpha]_3 &= -{1\over6}(\t\g_{mn})^{\alpha}(\cX_{X}\g^{m}\t)(\cX_{Y}\g^{n}\t)   \cr
[{\cal F}_{X,Y}^{mn}]_2 &=  - ({\cal X}_X\gamma^{[m}\theta)({\cal X}_Y\gamma^{n]}\theta) \ , 
}
$$
and further instances as to make the complete orders $\theta^{\leq 5}$ available are spelt out in appendix \HSstuff.
It is easy to see that these non-linear terms vanish in the single-particle case, and one
recovers the linearized expansions of \PolicastroVT.

Analogous theta-expansions for superfields \highmass\ of higher mass dimensions start with
\eqnn\THEmass
$$\eqalignno{
{\cal W}_P^{m\alpha}(x,\theta)&= e^{k_Px}\Big( \cX_P^{m\alpha} + {1\over4}(\theta\g_{np})^{\alpha}\cf^{m\vert np}_P+ \! \! \sum_{XY=P} \! \! \big[(\cX_{X}\g^{m}\theta)\cX^{\alpha}_Y-(X\leftrightarrow Y)\big]+ \ldots \Big) \cr
{\cal F}_P^{m|pq}(x,\theta)&=e^{k_Px}  \Big(  \cf_P^{m|pq} -(\cX^{m[p}_P\gamma^{q]}\theta)
+ \! \! \sum_{XY=P} \! \! \big[(\cX_{X}\gamma^{m}\theta)\cf^{pq}_Y-(X\leftrightarrow Y)\big] + \ldots \Big) \ ,&\THEmass\cr
}$$
where the lowest two orders $\sim \theta^2, \theta^3$ in the ellipsis
along with generalizations to higher mass dimensions are spelt out in appendix \HSstuff.

\subsec Combining HS gauge with BCJ gauge
\par\subseclab\HSandBCJ

\noindent
The steps in \toHS\ and \getL\ towards HS gauge can be literally repeated when starting with BCJ gauge:
\eqnn\BCJHS
$$\eqalignno{
\Bbb A^{{\rm BCJ-HS}}_\alpha &= \Bbb A^{{\rm BCJ}}_\alpha - [D_\alpha ,\Bbb L' ]
+ [\Bbb A^{{\rm BCJ}}_\alpha , \Bbb L'] &\BCJHS \cr
[{\cal D},\Bbb L' ]&= \theta^\alpha \Bbb A^{{\rm BCJ}}_\alpha
+ [\theta^\alpha \Bbb A^{{\rm BCJ}}_\alpha , \Bbb L'] \ .
}
$$
The multiparticle expansion of the gauge parameter $\Bbb L' $ can be constructed along the lines of \recL, 
where we again set $\Bbb L'(\theta=0) = 0$.
The resulting gauge combines the benefits of a simplified theta-expansion due to
\eqn\defBCJHS{
\theta^\alpha \Bbb A^{{\rm BCJ-HS}}_\alpha = 0
}
with a manifestation of the BCJ duality in cubic-diagram numerators
subject to Lie symmetries. The arguments of subsection \towHS\ give rise to theta-expansions
completely analogous to HS gauge, see \THEXone\ and appendix \HSstuff. The only difference is a redefinition of the
component Berends--Giele currents according to
\eqnn\newcompBG
$$\eqalignno{
\ce^m_P &\rightarrow {\cal A}_P^{m,{\rm BCJ}}(\theta=0)  = \ce^m_P + \sum_{XY=P} ( \ce^m_X \ch_Y - \ce^m_Y \ch_X) - k_P^m \ch_P
\cr
\cX^\alpha_P &\rightarrow {\cal W}^{\alpha,{\rm BCJ}}_P(\theta=0)  = \cX^\alpha_P + \sum_{XY=P} (\cX^\alpha_X \ch_Y - \cX^\alpha_Y \ch_X) \ ,
&\newcompBG
}$$
where the multiparticle gauge parameters contribute through their $\theta=0$ order,
\eqn\gaugepar{
\ch_{P} \equiv {\cal H}_P(\theta=0) \ .
}
The redefinitions in \newcompBG\ propagate to their counterparts at higher mass dimension via \compBGzero. Since
BCJ gauge already violates the Lorentz-gauge condition at the three-particle level, 
e.g. $k_m^{123} {\cal A}_{123}^{m,{\rm BCJ}} = - 2 s_{123} {\cal H}_{123}$, transversality
\transverseBG\ of the modified current $\ce^m_P \rightarrow {\cal A}_P^{m,{\rm BCJ}}(\theta=0) $ will no longer hold.

Similarly, the theta-expansions of higher-mass dimension Berends--Giele currents given in \THEmass\ and appendix \HSstuff\
preserve their structure after the replacements in \newcompBG. As mentioned
earlier, the BCJ
gauge appears naturally in the context of string amplitudes due to the redefinitions
induced by the double poles in OPE contractions. Hence, BCJ-HS gauge is particularly convenient for an 
accelerated approach to component amplitudes of the pure spinor superstring.

\newsec Application of Berends--Giele currents in Harnad--Shnider gauge

In this subsection, we sketch applications of multiparticle superfields in HS gauge
to scattering amplitudes in pure spinor superspace, relevant to both string and field theories. 
The identification of gluon and gluino components in supersymmetric kinematic factors is shown to simplify
enormously in HS gauge, in particular for large numbers of external legs.

\subsec Pure spinor superspace

Pure spinor superspace is obtained by supplementing ten-dimensional
superspace $\{x^m,\theta^\alpha\}$ with a bosonic Weyl spinor $\lambda^\alpha$ subject to the pure spinor constraint
\eqn\pscon{
(\l \gamma^m \l) = 0 \ .
}
Physical components in pure spinor superspace reside at the order $\l^3 \theta^5$ \psf, 
\eqn\lambdatheta{
\langle (\l \gamma^m \theta)  (\l \gamma^n \theta)  (\l \gamma^p \theta) (\theta \gamma_{mnp} \theta) \rangle = 2880 \ ,
}
and group theory fixes any other tensor structure in terms
of the above scalar \anomaly. The
prescription \lambdatheta\ guarantees that kinematic factors $S(\theta,\l)$ in
the cohomology of the BRST operator
\eqn\BRST{
Q \equiv \l^\alpha D_\alpha
}
yield supersymmetric and gauge invariant components $\langle S(\theta,\l) \rangle$ under the bracket \lambdatheta\ \psf.
On these grounds, various scattering amplitudes in ten-dimensional SYM have been
proposed by constructing BRST-invariant expressions with the
required propagator structure \refs{\towards, \nptFT, \MafraGJA, \MafraMJA}.
Also, cohomology arguments have given constructive input to the computation of
superstring amplitudes \refs{\nptTree, \MafraKH, \GreenBZA}.

Up to now, in order to extract the kinematic components from scattering amplitudes in pure spinor
superspace, the theta-expansions of the linearized superfields are
inserted into the recursive definitions of multiparticle superfields, 
leaving a huge number of tensor contractions of $\l^3 \theta^5$
for a computer-based evaluation \MafraPN.
Many kinematic factors obtained from this procedure 
have been gathered on the website \WWW.
HS gauge, on the other hand, drastically
reduces the number of different $\l^3 \theta^5$ contractions.
This makes kinematic factors with an arbitrary number of
external legs tractable for manual evaluation.

\subsec Applications at tree level

Tree-level kinematics of both the open superstring \nptTree\ and ten-dimensional SYM \nptFT\ can be expressed in terms of the building block
\eqn\treeMMM{
\langle M_A M_B M_C \rangle \ , \quad\ M_A \equiv \l^\alpha {\cal A}_\alpha^A \ .
}
BRST-invariant combinations of the building block \treeMMM\ descend from a
generating series of color-dressed tree-level amplitudes ${\cal M}^{{\rm SYM}}(1,2,\ldots,n)$ \MafraGIA,
\eqn\genser{
{1\over 3}{\rm Tr}\langle \Bbb V \Bbb V \Bbb V \rangle =
\sum_{n=3}^{\infty} (n-2) \sum_{i_1<i_2<\ldots<i_n} {\cal M}^{{\rm SYM}}(i_1,i_2,\ldots,i_n) \ , \ \ \ \ \ \ \Bbb V \equiv
\l^\alpha \Bbb A_\alpha \ .
} 
Since \genser\ is also invariant under non-linear
gauge transformations, the components of \treeMMM\ can be equivalently
evaluated in HS gauge for
arbitrary multiplicity,
\eqn\MMMresult{
\langle M^{{\rm HS}}_A M^{{\rm HS}}_B M^{{\rm HS}}_C \rangle =
{1\over 2} \ce_A^m \ce_B^n \cf_{mn}^C +  ({\cal X}_A \g_m {\cal X}_B) \ce_C^m
+ {\rm cyc}(A,B,C) \ .
}
The component currents $\ce_A^m, {\cal X}^\alpha_A$ and $\cf^{mn}_A$ defined in \sameBG\ and \moreBG\ 
can be obtained by truncating the superspace recursion \BGdef\ to \cFmn\ to $\theta=0$. By the theta-expansions 
in \THEXone, this component extraction involves no tensor structures $\sim \l^3 \theta^5$ other than
\eqnn\bose
$$\eqalignno{
\langle (\l  \g^m \t)  (\l \g^n  \t)
(\l  \g_r \t) (\t \g^{pqr} \t) \rangle  &=
32 (\delta^{mp}  \delta^{nq} -  \delta^{mq} \delta^{np}) & \bose \cr
\langle  (\l \g^m  \t )  (\l  \g^n  \t)
(\l  \g^p  \t)  (\g_n  \t)_\alpha  (\g_p  \t)_\beta  \rangle  &=
- 18 \g^m_{\alpha \beta} \ ,
}
$$
and elegantly settles the building blocks for components of tree-level
amplitudes. In a companion paper \otherpaper, it will be demonstrated
that \MMMresult\ reproduces the
Berends--Giele formula for bosonic tree amplitudes \BerendsME\ along with its
supersymmetric completion from the pure spinor superspace formula \nptFT.

The generating series \genser\ found appearance in \BerkovitsRB\ as a superspace
action for ten-dimensional SYM. The component evaluation in \MMMresult\ is
compatible with the component action of SYM in the sense that
\eqn\actions{
{1\over 3}{\rm Tr}\langle \Bbb V \Bbb V \Bbb V \rangle =  {\rm Tr} \Big( {1\over 4} \Bbb F_{mn} \Bbb F^{mn} + (\Bbb W  \gamma^m \nabla_m \Bbb W) \Big) \, \Big|_{\theta=0} \ .
}
The fermionic coupling vanishes on-shell by the Dirac equation
\DiracW\ and a total derivative $\partial_m$ has been discarded to relate 
\eqn\checkaction{
(\partial_m \Bbb A_n) \Bbb F^{mn} = \partial_m (\Bbb A_n \Bbb F^{mn}) - \Bbb A_n \big( [\Bbb A_m,\Bbb F^{mn}] + \gamma^n_{\alpha \beta} \{ \Bbb W^\alpha , \Bbb W^\beta \} \big)
}
through the expression for $\partial_m\Bbb F^{mn} $ in \cors.

\subsec Applications at loop level

In the same way as the building block \treeMMM\ is specific to tree amplitudes, any
loop order singles out specific scalar combinations of multiparticle superfields
which are BRST invariant at the linearized level, e.g.
\eqnn\oneloop
$$\eqalignno{
M_A (\l \g_m {\cal W}_B) (\l \g_n {\cal W}_C) {\cal F}^{mn}_D  \ &\leftrightarrow \
{\rm 1-loop} \ \refs{\MPS, \MafraKH, \EOMBBs}
&\oneloop \cr
(\l \g_{mnpqr} \l) (\l \g_s {\cal W}_A) {\cal F}^{mn}_B
{\cal F}^{pq}_C{\cal F}^{rs}_D   \ &\leftrightarrow \   {\rm 2-loop} \ \refs{\twoloop,\MafraMJA}
\cr
(\l \g_m {\cal W}^n_A) (\l \g_n {\cal W}^p_B) (\l \g_p {\cal W}^m_C)  \ &\leftrightarrow \   {\rm 3-loop} \ \MafraGIA \ .
}
$$
They describe the low-energy limit in string theory and are motivated by the
zero-mode saturation rules of the pure spinor formalism \refs{\psf,\MPS}. Moreover,
they are believed to represent box, double-box and Mercedes-star diagrams in SYM
amplitudes to arbitrary multiplicity, see \refs{\MafraGJA, \MafraMJA}. Again, HS
gauge as well as the theta-expansions in \THEXone, \THEmass\ and appendix~\HSstuff\ greatly simplify their
component evaluation via \lambdatheta.

In contrast to tree-level, loop amplitudes in SYM and superstring theory
additionally involve tensorial building blocks contracting the loop momenta where
HS gauge yields comparable benefits in the component evaluation. One-loop kinematic factors
generalizing \oneloop\ to arbitrary tensor rank have been constructed in
\cohomology, and some of them have been defined in terms of the superfields
$H_{12\ldots p}$ from the transformation to BCJ gauge. As will be described
elsewhere, kinematic factors with explicit reference to gauge parameters will
require extra care when adapted to different non-linear gauges. At any rate, HS gauge
for Berends--Giele currents sets new scales for the computational effort in component
evaluations.

\bigskip \noindent{\bf Acknowledgements:}
CRM wishes to acknowledge support
from NSF grant number PHY 1314311 and the Paul Dirac Fund.
We acknowledge support by the European
Research Council Advanced Grant No. 247252 of Michael Green. SL and OS are grateful 
to DAMTP in Cambridge for kind hospitality during various stages of this project,
and OS additionally thanks the IAS in Princeton for kind hospitality during
completion of this work. CRM is grateful to AEI in Potsdam
for the warm hospitality during intermediate stages of this work and
for partial financial support.

\appendix{A}{Proof of the Berends--Giele symmetries}
\applab\proofshuffle

\noindent
In this appendix, the symmetry property \shuff\ of Berends--Giele currents will be
proven from their recursive definition \BGdef. The idea is to regard the bracketing operation in
\eqn\BGproofA{
s_{AB} {\cal K}_{A \shuffle B} = \sum_{XY=A\shuffle B} {\cal K}_{[X,Y]}
}
as a linear and antisymmetric map ${\cal B}$ acting on a tensor product of words $X \otimes Y$,
\eqn\BGproofB{
{\cal B}: \ X \otimes Y \rightarrow  {\cal K}_{[X,Y]} \ , \ \ \ \ {\cal B}(X \otimes Y) = -{\cal B}(Y \otimes X) \ .
}
We will then show by induction that
\eqn\BGproofC{
s_{AB} {\cal K}_{A \shuffle B} = \sum_{XY=A\shuffle B}  {\cal B}(X \otimes Y) = 0 \ ,
}
starting with $0= {\cal K}_{1\shuffle 2}={\cal K}_{12}+{\cal K}_{ 21}$ by antisymmetry of the bracket.

As pointed out below \BGdef, the convention for deconcatenation sums $\sum_{XY=P}$
is to exclude the empty words $X=\emptyset$ and $Y=\emptyset$. Hence, they have to
be considered separately in relating \BGproofC\ to the deconcatenation coproduct for (possibly empty) words $P$,
\eqn\BGproofE{
\Delta(P) \equiv 1 \otimes P + P \otimes 1 + \sum_{XY=P} X \otimes Y \ .
}
This coproduct is known to be compatible with the shuffle product in the sense that
\eqnn\BGproofF
$$\eqalignno{
\Delta(A\shuffle B) &= \Delta(A) \shuffle \Delta(B) &\BGproofF
\cr
&= 1 \otimes (A\shuffle B) + (A\shuffle B) \otimes 1 + A\otimes B + B \otimes A
+  \sum_{PQ=A} \sum_{RS=B} (P \shuffle R) \otimes (Q \shuffle S)  \cr
& \ \ \ + \sum_{RS=B} \big( R \otimes (A\shuffle S) + (A \shuffle R) \otimes S \big)
+ \sum_{PQ=A} \big( P \otimes (Q\shuffle B) + (P \shuffle B) \otimes Q \big) \ ,
}
$$
see e.g.~section 1.5 in \ReutBook. The tensor product in \BGproofC\ can then be written as
\eqnn\BGproofG
$$\eqalignno{
&\sum_{XY=A\shuffle B}   X \otimes Y =  A\otimes B + B \otimes A
+  \sum_{PQ=A} \sum_{RS=B} (P \shuffle R) \otimes (Q \shuffle S) &\BGproofG \cr
& \ \ \ + \sum_{RS=B} \big( R \otimes (A\shuffle S) + (A \shuffle R) \otimes S \big)
+ \sum_{PQ=A} \big( P \otimes (Q\shuffle B) + (P \shuffle B) \otimes Q \big)   \ .
}
$$
In turns out that the right hand side is annihilated by ${\cal B}$ in \BGproofB\ since the first two 
terms $A\otimes B + B \otimes A$ drop out by antisymmetry of ${\cal B}$ and the remaining terms are 
mapped to the schematic form ${\cal K}_{[X\shuffle Y,Z]}$ under ${\cal
B}$ with all of $X,Y,Z \neq \emptyset$. By the bracketing rules \cAalpha\ to
\cWalpha, the latter yields antisymmetric combinations of ${\cal K}_{X\shuffle Y}$ and
${\cal K}_{Z}$ with $X\shuffle Y$ of multiplicity smaller than $|X|+|Y|+|Z|$.
Hence, we can set ${\cal K}_{X\shuffle Y} = 0$ by the inductive assumption which
concludes the proof of \BGproofC.

Note that the property $E_{A\shuffle B}=0$ can also be proved
similarly since the RHS of $E_P \equiv \sum_{XY=P} M_X M_Y$ is antisymmetric in
$X\leftrightarrow Y$. In addition,
the proof can be easily extended to ${\cal F}_P^{mn}$ and higher-mass dimension
superfields with recursive definition in \cFmn\ and \highmass: The deconcatenation
sums along with the non-linearities can be treated using the same arguments as
above, and the linear contributions from superfields of the same multiplicity
inherit the shuffle property of lower-mass dimension superfields.

\appendix{B}{BCJ gauge versus Lorentz gauge at rank five}
\applab\BCJapp

\ifig\figBGFive{The fourteen binary trees used in the definition of $\cK_{12345}$.}
{\epsfxsize=.89\hsize\epsfbox{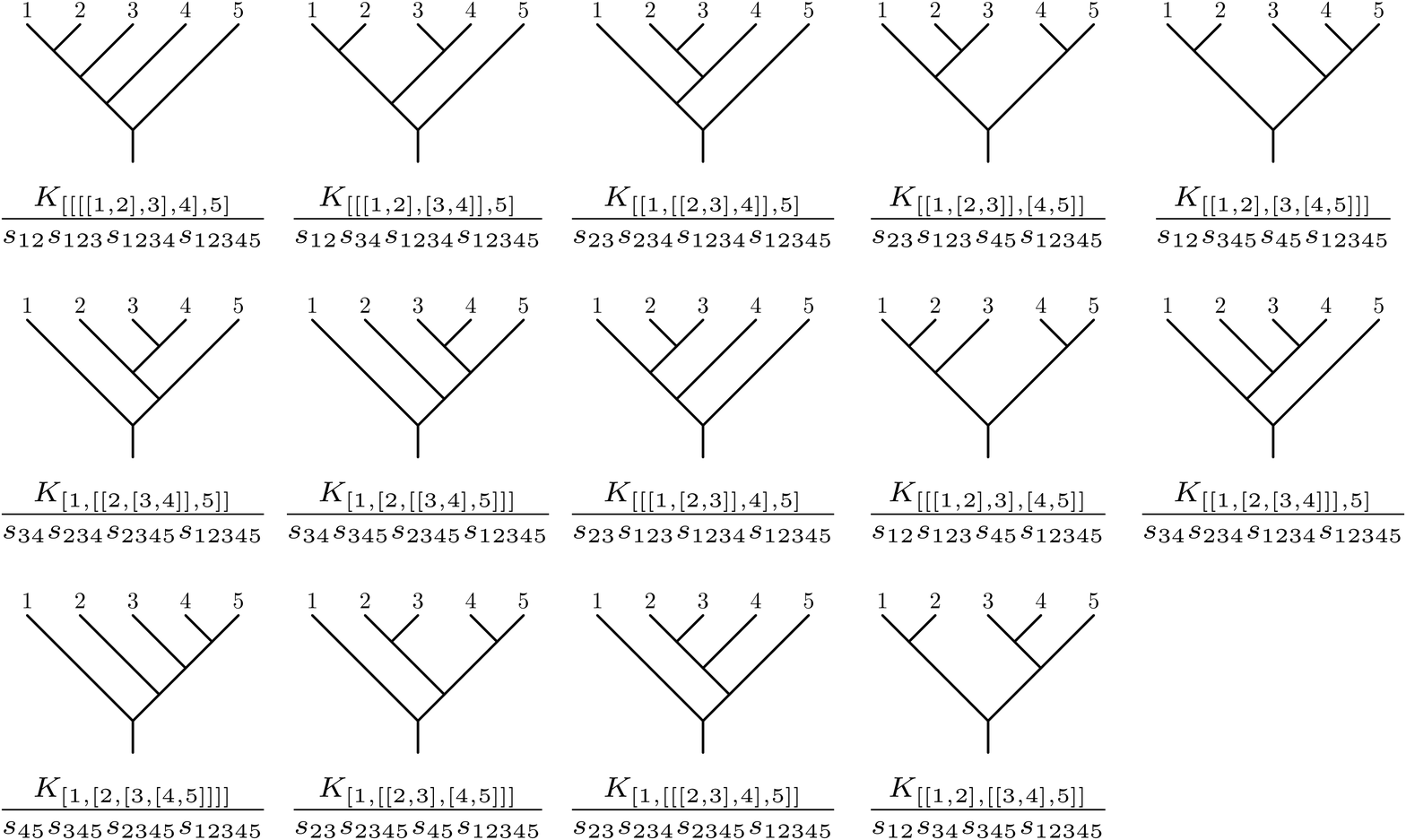}}

\noindent In this appendix, we verify that the supersymmetric Berends--Giele
currents at rank five in BCJ gauge and Lorentz gauge are related by a non-linear gauge transformation
as in \toshow. Straightforward but tedious calculations lead to
the following translation between local superfields in BCJ and Lorentz gauge,
\eqnn\BCJLorentzFive
\eqnn\BCJLorentzFivea
\eqnn\BCJLorentzFiveb
$$\eqalignno{
A_{[1234,5]}^{m} &=
\hat A^{m}_{[1234,5]}
 - k_{12345}^m \hat H_{[1234,5]}
         &\BCJLorentzFive\cr
&- (k^{1}\cdot k^{2}) (
        \hat H_{[134,5]} A_{2}^{m}
        + \hat H_{[14,5]} A_{23}^{m}
        + \hat H_{[13,5]} A_{24}^{m}
        + \hat H_{[13,4]} A_{25}^{m}
	- (1\leftrightarrow 2)
        )\cr
 &- (k^{12}\cdot k^{3}) (
        \hat H_{[124,5]} A_{3}^{m}
        + \hat H_{[12,5]} A_{34}^{m}
        + \hat H_{[12,4]} A_{35}^{m}
        - \hat H_{[34,5]} A_{12}^{m}
        )\cr
       & - (k^{123}\cdot k^{4}) (
        \hat H_{[123,5]} A_{4}^{m}
        + \hat H_{[12,3]} A_{45}^{m}
        )\cr
       & - (k^{1234}\cdot k^{5}) (
        \hat H_{[123,4]} A_{5}^{m}
        )\cr
A_{[123,45]}^{m} &=
        \hat A^{m}_{[123,45]}
        - k_{12345}^m
        \hat H_{[123,45]}\cr
       & - (k^{1}\cdot k^{2}) (
       \hat  H_{[13,45]} A_{2}^{m}
        + \hat  H_{[45,2]} A_{13}^{m}
	- (1\leftrightarrow 2))
&\BCJLorentzFivea
        \cr
        &- (k^{12}\cdot k^{3}) (
        \hat  H_{[12,45]} A_{3}^{m}
        +\hat  H_{[45,3]} A_{12}^{m}
        )\cr
        &- (k^{123}\cdot k^{45}) (
        \hat  H_{[12,3]} A_{45}^{m}
        )\cr
        &- (k^{4}\cdot k^{5}) (
         \hat H_{[123,4]} A_{5}^{m}
        - \hat H_{[123,5]} A_{4}^{m}
        )\cr
        A^{m}_{[[12,34],5]} &=
	\hat A^{m}_{[[12,34],5]}
        - k_{12345}^m
        \hat H_{[[12,34],5]} &\BCJLorentzFiveb
        \cr
&        - (k^{1}\cdot k^{2}) (
        \hat  H_{[34,2]} A_{15}^{m}
        - \hat  H_{[34,1]} A_{25}^{m}
        + \hat H_{[342,5]} A_{1}^{m}
        - \hat H_{[341,5]} A_{2}^{m}
        )\cr
        &- (k^{3}\cdot k^{4}) (
        \hat  H_{[12,3]} A_{45}^{m}
        - \hat  H_{[12,4]} A_{35}^{m}
        + \hat H_{[123,5]} A_{4}^{m}
        - \hat H_{[124,5]} A_{3}^{m}
        )\cr
        &- (k^{12}\cdot k^{34}) (
        \hat  H_{[12,5]} A_{34}^{m}
        - \hat  H_{[34,5]} A_{12}^{m}
        )\cr
        &- (k^{1234}\cdot k^{5}) (
        \hat  H_{[12,34]} A_{5}^{m}
        ) \ ,
}$$
where the second and third equations can be regarded as the definitions
of $\hat H_{[123,45]}$ and $\hat H_{[[12,34],5]}$. The solution of the
former is given in \topologies\ and \Hhatrecurs\ while the latter is
\eqn\crazytopology{
\hat H_{[[12,34],5]} =H_{[1234,5]} - H_{[1243,5]} -\half H_{[12,34]} (k_{1234} \cdot A_5) \ .
}
Plugging the above equations
into the generic definition of the rank-five
Berends--Giele current as displayed in \figBGFive, namely,
\eqnn\BGFiveKs
$$\eqalignno{
s_{12345}\cK_{12345} &=
	 {K_{[1,4532]}\over s_{2345} s_{345} s_{45}}
        - {K_{[1,3452]}\over s_{2345} s_{345} s_{34}}
        - {K_{[1,3425]}\over s_{2345} s_{234} s_{34}}
        + {K_{[1,2345]}\over s_{2345} s_{234} s_{23}}
	- {K_{[12,453]}\over s_{345} s_{12} s_{45}}\cr
      &\quad{}  + {K_{[12,345]}\over s_{345} s_{12} s_{34}}
        + {K_{[45,231]}\over s_{123} s_{23} s_{45}}
        - {K_{[45,123]}\over s_{123} s_{12} s_{45}}
	+ {K_{[3421,5]}\over s_{1234} s_{234} s_{34}}
        - {K_{[2341,5]}\over s_{1234} s_{234} s_{23}}\cr
     &\quad{}   - {K_{[2314,5]}\over s_{1234} s_{123} s_{23}}
        + {K_{[1234,5]}\over s_{1234} s_{123} s_{12}}
	+ {K_{[1,[23,45]]}\over s_{2345} s_{23} s_{45}}
	- {K_{[5,[12,34]]}\over s_{1234} s_{12} s_{34}}  \ ,&\BGFiveKs
}$$
leads to
\eqn\BGfivesol{
\cA^{m,\rm BCJ}_{12345} = \cA^{m,\rm L}_{12345} - k_{12345}^m \cH_{12345}
+ \cA^m_1 \cH_{2345} + \cA^m_{12} \cH_{345}
- \cA^m_5 \cH_{1234} - \cA^m_{45} \cH_{123} \ .
}
By the vanishing of ${\cal H}_i$ and ${\cal H}_{ij}$, this reproduces the non-linear gauge transformation \toshow\ 
at multiplicity five.


\appendix{C}{Theta-expansions in Harnad--Shnider gauge}
\applab\HSstuff

\subsec Theta-expansions of ${\cal A}^P_\alpha,{\cal A}_P^m,{\cal W}_P^\alpha, {\cal F}_P^{mn}$

The component prescription \lambdatheta\ in pure spinor superspace
requires the theta-expansion of the enclosed superfields up to the order $\theta^5$.
The expansions up to $\theta^5$ of the Berends--Giele
currents ${\cal A}^P_\alpha,{\cal A}_P^m,{\cal W}_P^\alpha, {\cal F}_P^{mn}$
in HS gauge can be found in \THEXone\ up to deconcatenation terms. These are now
spelt out:
\eqnn\appCa
\eqnn\appCb
\eqnn\appCc
\eqnn\appCd
\eqnn\appCe
\eqnn\appCf
$$\eqalignno{
[{\cal A}_{X,Y}^m]_5 =&\pplus
{1\over320}(\t\g^{mnr}\t)(\t\g_{rpq}\t)({\cal X}_{X}\g_{n}\t)\cf^{pq}_Y -(X\leftrightarrow Y) &\appCa \cr
[{\cal W}_{X,Y}^\alpha]_4 =&
-{1\over64}(\t\g^{\phantom{m}q}_{m})^{\alpha}(\t\g_{qnp}\t)({\cal X}_{X}\g^{m}\t)\cf^{np}_Y-(X\leftrightarrow Y) &\appCb \cr
[{\cal W}_{X,Y}^\alpha]_5=&\pplus {1\over120}(\t\g^{\phantom{m}q}_{m})^{\alpha}(\t\g_{npq}\t)(\cX_{X}\g^{m}\t)(\cX^{n}_{Y}\g^p\t)\cr
&+ {1\over240} (\t\g^{\phantom{n}q}_{n})^{\alpha}(\t\g_{mpq}\t)(\cX_{X}\g^{m}\t)(\cX^{n}_{Y}\g^p\t)\cr
&-{1\over1280}(\t\g^{rs})^\alpha(\t\g_{mnr}\t)(\t\g_{pqs}\t)\cf^{mn}_{X}\cf^{pq}_{Y}
-(X\leftrightarrow Y)  &\appCc\cr
[{\cal F}_{X,Y}^{mn}]_3=&\pplus {1\over8}    (\theta\gamma_{pq}^{\phantom{pq}[m}\theta)({\cal X}_{X}\gamma^{n]}\theta)\cf^{pq}_Y-(X\leftrightarrow  Y)     &\appCd \cr
[{\cal F}_{X,Y}^{mn}]_4=
&-{1\over12}(\t\g^{\phantom{pq}[m}_{pq}\t)(\cX_{X}\g^{n]}\t)(\cX^{p}_{Y}\g^{q}\t)\cr
&{}-{1\over24}(\t\g^{pq[m}\t)(\cX_{X}\g_{p}\t)(\cX^{n]}_{Y}\g_{q}\t)\cr
&{}-{1\over128}(\t\g^{[m}_{\phantom{[m}pq}\t)(\t\g^{n]}_{\phantom{n]}rs}\t)\cf^{pq}_{X}\cf^{rs}_{Y}
-(X\leftrightarrow Y) &\appCe \cr
[{\cal F}_{X,Y}^{mn}]_5 =
&-{1\over192}(\t\g^{[m}_{\phantom{m}ps}\t)(\cX_{X}\g^{n]}\t)\cf^{p\vert qr}_{Y}(\t\g^{s}_{\phantom{s}qr}\t) \cr
& -{1\over320}(\cX_{X}\g^{p}\t)(\t\g^{\phantom{ps}[m}_{ps}\t)\cf^{n]\vert qr}_{Y}(\t\g^{s}_{\phantom{s}qr}\t)\cr
& -{1\over320}(\t\g^{\phantom{ps}[m}_{ps}\t)(\cX^{n]}_{X}\g^{p}\t)\cf^{qr}_Y(\t\g^{s}_{\phantom{s}qr}\t)\cr
& +{1\over96}(\t\g^{[m}_{\phantom{[m}pq}\t)(\t\g^{n]}_{\phantom{n]}rs}\t)(\cX^{p}_{X}\g^{q}\t)\cf^{rs}_Y
-(X\leftrightarrow Y)\cr
[{\cal F}_{X,Y,Z}^{mn}]_5=&
- {1\over24} (\t\g^{\phantom{pq}[m}_{pq}\t)(\cX_{X}\g^{n]}\t)(\cX_{Y}\g^{p}\t)(\cX_{Z}\g^q\t)   + (X \leftrightarrow Z)  \ . &\appCf
\cr
}$$

\subsec Theta-expansions of the simplest higher-mass dimension superfields

For the simplest superfields of higher mass dimension, the theta-expansion in HS gauge that starts as in \THEmass\ and has the following second and third order:
\eqnn\moreTHEX
$$\eqalignno{
[{\cal W}_P^{m\alpha}]_2 &= -{1\over4} (\t\g_{np})^{\alpha}({\cal X}^{mn}_{P}\g^{p}\t)
+\sum_{XY=P}\big[ {1\over 4}(\t\g_{np})^{\alpha}({\cal X}_X\g^m\t)\cf_Y^{np}&\cr
&\ \ \ -{1\over8}(\t\g^{m}_{\phantom{m}np}\t){\cal X}^{\alpha}_X\cf_Y^{np}
-(X\leftrightarrow Y)\big]& \cr
[{\cal W}_P^{m\alpha}]_3 &= -{1\over48}(\t\g^{\phantom{n}r}_{n})^{\alpha}(\t\g_{rpq}\t)\cf^{mn\vert pq}_{P}
+\sum_{XY=P}\big[ -{1\over4} (\t\g_{np})^{\alpha}({\cal X}_{X}\g^{m}\t)({\cal X}^{n}_{Y}\g^{p}\t)&\cr
&\ \ \ -{1\over6}(\t\g_{np})^{\alpha}({\cal X}_{X}\g^{n}\t)({\cal X}^{m}_{Y}\g^{p}\t)
-{1\over12}(\t\g^{m}_{\phantom{m}np}\t)({\cal X}^{n}_{X}\g^{p}\t){\cal X}_Y^{\alpha}&\cr
&\ \ \ -{1\over32}(\t\g_{np})^{\alpha}(\t\g^{m}_{\phantom{m}qr}\t)\cf^{np}_X\cf^{qr}_Y-(X\leftrightarrow Y)   \big]&\cr
%
%
[{\cal F}_P^{m|pq}]_2&=  -{1\over8}   \cf^{m[p}_{\phantom{m[p}\vert nr}(\t\g^{q]nr}\t)-\sum_{XY=P}\big[({\cal X}_X\g^m\t)({\cal X}_Y^{[p}\g^{q]}\t) &\moreTHEX \cr
&\ \ \ 
+({\cal X}^{m}_{X}\g^{[p}\t)({\cal X}_{Y}\g^{q]}\t)
+{1\over 8}(\t\g^{m}_{\phantom{m}nr}\t)\cf_X^{pq}\cf_Y^{nr}-(X\leftrightarrow Y)   \big]&\cr
[{\cal F}_P^{m|pq}]_3&= 
{1\over12} ({\cal X}^{m[p}_{B\phantom{[m}n}\g_{r}\t)(\t\g^{q]nr}\t)
+\sum_{XY=P}\big[{1\over8}({\cal X}_{X}\g^{m}\t)(\t\g^{[p}_{\phantom{[p}nr}\t)\cf^{q]\vert nr}&\cr
&\ \ \ +{1\over8}(\t\g^{\phantom{nr}[p}_{nr}\t)({\cal X}_{X}\g^{q]}\t)\cf^{m\vert nr}_{Y}
-{1\over8}({\cal X}^{m}_{X}\g^{[p}\t)(\t\g^{q]}_{\phantom{q]}nr}\t)\cf^{nr}_Y&\cr
&\ \ \ +{1\over8}(\t\g^{m}_{\phantom{m}nr}\t)({\cal X}^{[p}_{X}\g^{q]}\t)\cf^{nr}_Y
-{1\over 12}(\t\g^{m}_{\phantom{m}nr}\t)({\cal X}^{n}_{X}\g^{r}\t)\cf^{pq}_Y -(X\leftrightarrow Y)   \big]&\cr
&\ \ \ +\sum_{XYZ=P}\big[({\cal X}_{X}\g^{[p}\t)({\cal X}_{Y}\g^{q]}\t)({\cal X}_{Z}\g^{m}\t) 
+(X\leftrightarrow Z)   \big]
\ . \cr
%
}
$$

\subsec Theta-expansions of generic higher-mass dimension superfields

For superfields of higher mass dimension as defined in \highmass, the theta-expansion in HS gauge is governed by the recursion
\eqnn\mosttheta
$$\eqalignno{
[\Bbb W^{N\alpha}]_k&={1\over k}\bigg\{{1\over4}(\t\g_{pq})^\alpha[\Bbb F^{N\vert pq}]_{k-1}
+ \! \! \! \!\sum_{M\in P(N)\atop M\ne0}\sum_{l=0}^{k-1} \big\{   \, ([\Bbb W]_l\g\t)^M \, ,
\, [\Bbb W^{(N\setminus M)\alpha}]_{k-l-1}  \, \big\}\bigg\}&\cr
[\Bbb F^{N\vert pq}]_k&=-{1\over k}\bigg\{([\Bbb W^{N[p}]_{k-1}\g^{q]}\t)
- \! \! \!\! \sum_{M\in P(N)\atop M\ne0}\sum_{l=0}^{k-1} \big[ \, ([\Bbb W]_l\g\t)^M \, ,
\, [\Bbb F^{(N\setminus M)\vert pq}]_{k-l-1} \, \big] \bigg\}\ .&\mosttheta
}
$$
We are using multi-index notation $N\equiv n_1n_2\ldots n_{k}$, where the power set $P(N)$
consists of the $2^{k}$ ordered subsets of $N$, and
$(\Bbb W \gamma)^N\equiv (\Bbb W^{n_1\ldots n_{k-1}} \gamma^{n_{k}})$.
Their resulting theta-expansion to subleading order is given by
\eqnn\mostTHEX
$$\eqalignno{
{\cal W}_P^{N\alpha}(\t)&= {\cal X}_P^{N\alpha}+{1\over 4}(\t\gamma_{pq})^\alpha \cf^{N|pq}_{P}\cr
&\ \ \ +\sum_{XY=P}\sum_{M\in P(N)\atop M\ne0} \big[ ( {\cal X}^{\ }_X\gamma\t)^M {\cal X}^{(N\setminus M)\alpha}_Y-({\cal X}^{\ }_Y\gamma\t)^M {\cal X}^{(N\setminus M)\alpha}_X \big] + \ldots \cr
{\cal F}^{N|pq}_P(\t)&= \cf^{N|pq}_P-( {\cal X}^{N[p}\gamma^{q]}\t) &\mostTHEX \cr
&\ \ \ +\sum_{XY=P}\sum_{M\in P(N)\atop M\ne0} \big[( {\cal X}^{\ }_X\gamma\t)^M \cf^{(N\setminus M)\vert pq}_Y-({\cal X}^{\ }_Y\gamma\t)^M \cf^{(N\setminus M)\vert pq}_X \big] + \ldots \ .
}
$$

\listrefs

\bye